\documentclass[12pt]{article}
\pdfoutput=1
\usepackage{graphics, color}
\usepackage{graphicx}
\usepackage{amsmath}
\usepackage{amssymb}
\usepackage{amsfonts}
\usepackage{tensor}
\usepackage[usenames,dvipsnames]{xcolor}
\usepackage[normalem]{ulem}
\usepackage[bottom]{footmisc}

\definecolor{orange}{rgb}{1,0.5,0}
\definecolor{grey}{rgb}{.5,.5,.5}
\definecolor{bluegreen}{rgb}{0,.5,.5}
\definecolor{darkgreen}{rgb}{0,.5,0}

\newcommand{\sect}[1]{\section{#1}\setcounter{equation}{0}}

\def\gsim{\, \rlap{$>$}{\lower 1.1ex\hbox{$\sim$}}\,}
\def\lsim{\, \rlap{$<$}{\lower 1.1ex\hbox{$\sim$}}\,}
\newcommand{\be}{\begin{equation}}
\newcommand{\ee}{\end{equation}}
\newcommand{\bea}{\begin{eqnarray}}
\newcommand{\eea}{\end{eqnarray}}

\newcommand{\bsig}{\boldsymbol\sigma}
\newcommand{\bchi}{\boldsymbol\chi}
\newcommand{\bB}{\boldsymbol B}
\newcommand{\bracedot}{\cdot}
\newcommand{\transd}{d}

\begin{document}


\begin{titlepage}
\bigskip
\bigskip\bigskip\bigskip
\centerline{\Large \bf Higher Spin Holography,}
\centerline{\Large \bf   RG,  and  the Light Cone}

\bigskip\bigskip\bigskip
\bigskip\bigskip\bigskip

 \centerline{{\bf Eric Mintun,}\footnote{\tt mintun@physics.ucsb.edu}*
 {\bf Joseph Polchinski}\footnote{\tt joep@kitp.ucsb.edu }*${}^\dagger$}
 \bigskip
\centerline{\em *Department of Physics}
\centerline{\em University of California}
\centerline{\em Santa Barbara, CA 93106 USA}
\bigskip
\centerline{\em ${}^\dagger$Kavli Institute for Theoretical Physics}
\centerline{\em University of California}
\centerline{\em Santa Barbara, CA 93106-4030 USA}

\bigskip\bigskip\bigskip

\begin{abstract}
We revisit the derivation of  higher spin bulk theory using the renormalization group in the dual field theory.  We argue that existing proposals have problems already at the level of linearized perturbations on \mbox{AdS}.  This is due to the form of the cutoff, which must act on bilinears in the fundamental fields rather than on the fields themselves.  For the light-cone collective field theory, we show that the RG produces the correct linearized perturbations.  We relate this to the precursor formula of de Mello Koch, Jevicki, Jin and Rodrigues, and we also elaborate on that result.  The covariant RG and bulk interactions remain problems for the future.

\end{abstract}
\end{titlepage}

\baselineskip = 16pt

\baselineskip = 16pt

\setcounter{footnote}{0}

\tableofcontents

\section{Introduction}

The emergence of spacetime in gauge-gravity duality is one of the central principles in quantum gravity, but the details remain mysterious.  It is often said that the AdS radial direction $z$ emerges from the energy or renormalization group scale of the dual field theory, but it is not clear whether this correspondence can be made sharp.  The AdS coordinate $z^{-1}$ and the CFT energy do scale in the same way under dilatations~\cite{Maldacena:1997re}.  However, for any given process there is an $O(1)$ uncertainty, or even more, in what we mean be the characteristic energy.  Then $\delta z/z \sim \delta E/E = O(1)$ translates into an uncertainty of order $R_{\rm AdS}$ in the corresponding radial position, parametrically larger than the Planck or string scale on which one expects local physics.

The question we would like to investigate here is whether there is some RG cutoff on the field theory that is dual to a sharp cutoff on $z$.  The idea of relating such cutoffs was used in Ref.~\cite{Maldacena:1996ds}, but only in an order-of-magnitude sense.  There has been a large literature developing the parallel between the supergravity equations and the renormalization group, but without reference to a specific cutoff on the QFT.

If one could find a such a cutoff, one could then hope to derive the bulk theory from the RG.  The approach~\cite{Lee:2009ij,Heemskerk:2010hk}  indicates how the bulk fields can emerge in this way.  However, in its current form this does not lead to a local theory in the bulk.

In this paper we explore this question for the higher-spin/vector model duality~\cite{Sundborg:2000wp,Sezgin:2002rt,Klebanov:2002ja}.  This has effectively been derived in Refs.~\cite{Maldacena:2011jn}, by showing that the symmetries of the bulk theory fully determine the dual \mbox{QFT}.  However, we are interested in going the other way, starting with the QFT and seeing how to identify the local structure.  For this purpose the RG may be a more robust tool, and not so dependent on a large symmetry algebra.  

In this theory there is not a hierarchy of operator dimensions, so one does not really expect local physics in the bulk.  Nevertheless, the Vasiliev higher spin theory~\cite{Vasiliev:1999ba,Bekaert:2005vh}  has local field equations, and so provides a target for an RG derivation.
This problem has been investigated in Refs.~\cite{Douglas:2010rc,Sachs:2013pca,Leigh:2014tza}, which identify parallels between the structure of the RG and the structure of Vasiliev's  theory.  However, we believe that these cannot be complete, because of the form of the cutoff assumed.

The $U(N)$ or $O(N)$ quantum numbers are not visible in the bulk, so the bulk cutoff necessarily acts on singlet fields.  In order to implement this, one would need to impose the QFT cutoff on invariant bilinear or single-trace fields (for a vector or matrix theory, respectively).  Indeed, the best guess as to the nature of $z$ is that it corresponds to the size of a $U(N)$ or $O(N)$ singlet state in the bulk.\footnote{Another approach uses conformal perturbation theory~\cite{Bzowski:2012ih}.}
Refs.~\cite{Douglas:2010rc,Sachs:2013pca,Leigh:2014tza} use instead standard cutoffs on the momenta of the $U(N)$ or $O(N)$ vectors.  The most immediate problem here is that the latter cutoff leads only to a single-bilinear action and so to a first-order bulk equation.   Thus even at the level of perturbations of AdS, one does not obtain the correct  bulk equations of motion or bulk-to-boundary propagator.  

Most of our paper is thus directed at this lowest order in $1/N$, free fields in the bulk.
For reasons that we review in \S\ref{HRG}, a cutoff on the bilinear allows for a second-order field equation~\cite{Heemskerk:2010hk,Lee:2009ij}.  We have tried to use this clue to obtain an RG with the correct bulk propagator, but thus far have not found a natural solution.  However, in light-cone frame the answer already exists in the literature, in the collective field approach of Koch, Jevicki, Jin, and Rodrigues~\cite{Koch:2010cy}.   Brodsky and de Teramond have given a closely related treatment of QCD~\cite{Brodsky:2006uqa}.  Ref.~\cite{Koch:2010cy} gives, to leading order in $1/N$,  a simple and explicit expression for the precursors, the QFT operators that map to local fields in the bulk, with the latter being described by the light cone bulk analysis of Metsaev~\cite{Metsaev:1999ui}. This can be reinterpreted in RG language.

In \S\ref{HSLC} we write the CFT in terms of bilocal fields on the light cone.  We then show that a cutoff on the separation of the bilinear leads to a second order RG equation, and we verify that this is the same as the light-cone form~\cite{Metsaev:1999ui} of the Fronsdal equation~\cite{Fronsdal:1978rb}.  We work out in detail the mapping between RG and bulk fields in terms of their $SO(D-1)$ spins.

In \S\ref{HSP}, we relate this discussion to the precursor language~\cite{Koch:2010cy}.  We fill in some details of the holographic dictionary in light-cone gauge, and its relation to the de Donder form.  In \S\ref{Conclusion} we discuss some issues with extending this to higher order in $1/N$.

\sect{Holographic renormalization group}
\label{HRG}
\subsection{Review}
\label{HRGReview}
We begin by reviewing the formalism of~\cite{Heemskerk:2010hk}; similar ideas are explored in~\cite{Lee:2009ij,Faulkner:2010jy}.  Split the bulk path integral into UV and IR pieces at radius $z=\ell$, 
\bea
Z_{\mathrm{B}} &=& \int \mathcal{D} \Phi|_{z > \ell} \,\mathcal{D} \tilde{\Phi}\, \mathcal{D} \Phi|_{z<\ell} \, e^{-S_{\mathrm{B}}|_{z>\ell} - S_{\mathrm{B}}|_{z<\ell}}  \label{zsep1}
\nonumber\\
&\equiv& \int \mathcal{D} \tilde{\Phi}\, \Psi_{\rm IR}(\ell, \tilde{\Phi} ) \Psi_{\rm UV} (\ell, \tilde{\Phi} ) \label{zsep2}\, ,
\eea
where $\tilde{\Phi^i} = \Phi^i|_{z=\ell}$.  We use B to denote the bulk theory.  The two halves of the path integral each satisfy a radial evolution equation of Hamiltonian form,
\bea
\partial_\ell \Psi_{\rm IR} (\ell, \tilde{\Phi}) &=& H (\tilde{\Phi}, \tilde{\Pi} ) \Psi_{\rm IR}(\ell, \tilde{\Phi}) \nonumber \, ,\\
\partial_\ell \Psi_{\rm UV} (\ell, \tilde{\Phi}) &=& -H (\tilde{\Phi}, -\tilde{\Pi} ) \Psi_{\rm UV}(\ell, \tilde{\Phi}) \label{ham} \,,
\eea
where $\tilde\Pi_i = -i \delta/\delta\tilde\Phi^i$.  Together these imply that the path integral is independent of the separation radius $\ell$.

We would like to ask whether there is some cutoff in the CFT such that
\be
\Psi_{\rm IR}(\ell, \tilde{\Phi}^i) = \int \mathcal{D} \phi \exp \left [ -S_{0}(\ell, \phi) + \int d^d x \,\tilde{\Phi}^i(x) \mathcal{O}_i(x) \right ] \, . \label{dict}
\ee
Here the UV cutoff $\ell^{-1}$ has been built into the action $S_0$.  The $\phi$ stand for generic $N$-matrix or $N$-vector fields, and the $\mathcal{O}_i$ are correspondingly some complete set of local single-trace or bilinear operators.  The data on the two sides have a natural match, between the interface value of the fields and the couplings of the single-trace operators.\footnote{To define the radial separation~(\ref{zsep1}) it is necessary to choose coordinates, and presumably to fix the other bulk gauge invariances as well.  The mapping will depend on these choices.}
Inserting the dictionary~(\ref{dict}) into the bulk path integral~(\ref{zsep2}) gives
\bea
Z_{\mathrm{B}} &=& \int \mathcal{D} \phi \,e^{ -S_{0}(\ell, \phi)} \int  \mathcal{D} \tilde{\Phi}\, \Psi_{\rm UV} (\ell, \tilde{\Phi} )  e^{ \int d^d x \,\tilde{\Phi}^i(x) \mathcal{O}_i(x)}   \nonumber\\
&\equiv&  \int \mathcal{D} \phi \,e^{ -S_{0}(\ell, \phi) - S_1(\ell, \phi) } \,.   \label{legendre}
\eea

The Hamiltonian equations~(\ref{ham}) now become RG equations for $\Psi_{\rm IR}$ and for the Wilsonian interaction $S_1$.  The latter is
\be
\dot {(e^{-S_1})} = - H(-\delta/\delta {\cal O}, i {\cal O}) e^{-S_1} \,. \label{uvrg}
\ee
Equivalently, an RG equation can be reinterpreted as a bulk field equation.  The expectation value of a bulk field is
\be
\langle \Psi_{\rm UV}(\ell) | \tilde\Phi^i(x) | \Psi_{\rm IR}(\ell) \rangle = \int {\cal D}  \tilde\Phi\, \Psi_{\rm UV}(\ell,\tilde\Phi) \tilde\Phi ^i(x)\Psi_{\rm IR}(\ell,\tilde\Phi) \,.
\ee
It evolves as
\be
\partial_\ell  \langle \Psi_{\rm UV}(\ell) | \tilde\Phi^i(x) | \Psi_{\rm IR}(\ell) \rangle
= -\langle \Psi_{\rm UV}(\ell) | [H, \tilde\Phi^i(x)] | \Psi_{\rm IR}(\ell) \rangle \,. \label{heis}
\ee
Given an RG equation we can in this way derive the corresponding bulk field equation.
However, this will in general be nonlocal on a scale of order $R_{\rm AdS}$~\cite{Heemskerk:2009pn}, and the challenge is to see the local physics.\footnote{In Ref.~\cite{Heemskerk:2009pn} it was shown in one example that a large gap in the operator spectrum leads to locality down to the string scale.  That argument depended on an orgy of hypergeometric functions, thought it has been simplified using the Mellin representation~\cite{Penedones:2010ue}.  One might hope that the RG would be a more robust and physical tool for demonstrating this, but so far this has not been realized.}

We have written the RG for a Euclidean spacetime, but the discussion applies as well to the Lorentzian case by inserting $i$'s.  As emphasized in Ref.~\cite{Balasubramanian:2012hb}, $\Psi_{\rm UV} $ must be nonlocal in time, because it included effects of excitations that propagate into the UV region and return at some later time.  This is true in both the bulk and CFT pictures.  However, we see from the bulk point of view that the radial $H$, rather than $\Psi_{\rm UV} $, can still be local down to the string scale: this is where the locality of the theory is encoded.

\subsection{Vector theories}
\label{HRGVector}
We now apply this to the theory of a $U(N)$ vector $\phi^a$; the treatment of a real $O(N)$ vector is parallel.  In this subsection we consider the usual momentum cutoff on $\phi^a$~\cite{Douglas:2010rc,Sachs:2013pca,Leigh:2014tza},
\be
S_0(\ell) = \int \frac{d^dp}{(2\pi)^d}\, p^2 K^{-1}(\ell^2 p^2) \phi^\dagger(p) \!\bracedot\! \phi(-p) \,. \label{ordcut}
\ee
The complete set of invariants ${\cal O}_i(x)$ become the invariant bilinears $\phi^\dagger(x) \!\bracedot\! \phi(y)$.  The bulk fields $\tilde\Phi^i(x)$ become bilinears $B(x,y)$; for convenience we leave off the tilde henceforth, and also we use the same notation for fields and their Fourier transforms.  Thus
\be
\Psi_{\rm IR}(\ell,B) = \int {\cal D}\phi \, \exp \left\{-S_0(\ell) +  \int dp \,dq\,B(-q,-p) \phi^\dagger(p) \!\bracedot\! \phi(q) \right\} \,. \ee

One has immediately the differential equation
\begin{eqnarray}
\dot \Psi_{\rm IR}(\ell,B) &=&
-\int {\cal D}\phi \, e^{-S_0 + \int B \psi} \int \frac{d^dp}{(2\pi)^d}\, p^2\dot K^{-1}(\ell^2 p^2)   \phi^\dagger(p) \!\bracedot\! \phi(-p) \nonumber\\
&=&
-\int \frac{d^dp}{(2\pi)^d}\, p^2 \dot K^{-1}(\ell^2 p^2) \frac{\delta}{\delta B(p,-p)} \Psi_{\rm IR}(\ell,B) \,. \label{rgtriv}
\end{eqnarray}
where a dot denotes $\ell \partial_\ell$.  By itself this has no content.  To convert it to an RG equation we use also the Schwinger-Dyson equation
\begin{eqnarray}
0 &=& \int {\cal D}\phi\, \partial_{\phi(-p)} \!\bracedot\! \left( \phi(q) e^{-S_0 + \int B \phi^\dagger\phi} \right) \,,
\end{eqnarray}
which implies
\begin{eqnarray}
\frac{\delta}{ \delta B(p,q)} \Psi_{\rm IR}(\ell,B) &=& \frac{ K(\ell^2 p^2)}{p^2} \left[ N (2\pi)^d \delta^d(p+q) + \int \frac{d^ds}{(2\pi)^d}\, B(-p,s) \frac{\delta}{\delta B(q,s)} \right]\Psi_{\rm IR}(\ell,B) \nonumber\\
&=& \frac{ K(\ell^2 q^2)}{q^2} \left[ N (2\pi)^d \delta^d(p+q) + \int \frac{d^ds}{(2\pi)^d}\, B(s,-q)  \frac{\delta}{\delta B(p,s)} \right] \Psi_{\rm IR}(\ell,B) 
\,,\nonumber\\ \label{sde}
\end{eqnarray}
where the second line is obtained by conjugating the first.  Applying this twice to the differential equation~(\ref{rgtriv}), one obtains
\begin{eqnarray}
\dot \Psi_{\rm IR} &=& -\left[ N \delta(0) 
\int \frac{d^dp}{(2\pi)^d}\, K \dot K^{-1}+ \int \frac{d^dp}{(2\pi)^d}\,\frac{d^ds}{(2\pi)^d} \,  K\dot K^{-1} B(p,q) \frac{\delta}{\delta B(p,q)} \right] \Psi_{\rm IR}  \nonumber\\
&=& \int \frac{d^dp}{(2\pi)^d} \left[ N \delta(0) \dot K  K^{-1}
 + N  \frac{\dot K}{p^2}  B(p,-p) + \int \frac{d^dq}{(2\pi)^d}\,\frac{d^dr}{(2\pi)^d}\, \frac{\dot K}{p^2} B(q,p) B(-p,r) \frac{\delta}{\delta B(q,r)}\right] \Psi_{\rm IR} \,;
 \nonumber\\ \label{rg1}
\end{eqnarray}
(the argument of $K$ is everywhere $\ell^2 p^2$). The $\delta(0)$ term can be absorbed into the path integral measure, and we drop it henceforth.

This is now the Wilson renormalization group, in the form~\cite{Polchinski:1983gv}.  We can generalize by including a
field definition $\dot B(p,q) = v(p,q,B)$, which adds a multiple of the Schwinger-Dyson equation~(\ref{sde}) to the RG:
\begin{eqnarray}
\dot \Psi_{\rm IR} 
&=& \int \frac{d^dp}{(2\pi)^d} \Biggl[ \frac{N}{p^2}  \left({\dot K} B(p,-p) + K v(p,-p, B)\right) - \int dq\, v(p,q,B) \frac{\delta}{ \delta B(p,q)}
\nonumber\\
&& \qquad +  \int \frac{d^dq}{(2\pi)^d}\,\frac{d^dr}{(2\pi)^d}\,  \left({\dot K} B(p,q) + K v(p,q, B)\right)
 \frac{1}{p^2} B(r,-p)  \frac{\delta}{ \delta B(r,q)}\Biggr] \Psi_{\rm IR} \,.\nonumber\\
\end{eqnarray}

This translates into a radial Hamiltonian $H$ as in Eq.~(\ref{ham}).  We see that $H$ will contain only terms of zeroth and first order in $\Pi_B$.  The field equation~(\ref{heis}) for $B$ is then first order in derivatives.  In particular, linearizing in $B$ gives simply
\be
\dot B(p,q) =- \int \frac{d^dr}{(2\pi)^d}\,\frac{d^ds}{(2\pi)^d}\, B(r,s) \frac{\delta v(p,q,B)}{\delta B(r,s)} \bigg|_{B=0} \,. \label{beq}
\ee
To obtain a second order equation for $B$, we would need a $\Pi^2$ term in $H$.  The scale derivative of $B$ would then involve $\Pi$, and vice versa, producing a second order equation.   However, we do not get such a term from this regulator.\footnote{In the Vasiliev form of higher-spin theory~\cite{Vasiliev:1999ba,Bekaert:2005vh}, there is a multicomponent first order equation, which becomes effectively second order due to a tracelessness condition.  It does not seem that this can happen here.}   
We see from the RG~(\ref{uvrg}) that $\Pi^2$ terms are associated with the flow of double-trace terms.  However, for $N$-vector fields with the standard cutoff~(\ref{ordcut}), double-bilinear interactions are not generated.

For another perspective, let us note that we can derive the RG equation~(\ref{rg1}) more directly in the case of a sharp cutoff,
\be
K = 1,\ p^2 \ell^2< 1 \,, \quad K = 0,\ p^2 \ell^2 > 1 \,.
\ee
Doing the gaussian integral for the shell $(\ell + d\ell)^{-2} < p^2 < \ell^{-2}$ immediately gives ~(\ref{rg1}), with $\dot K
= -2 \delta(p^2 \ell^2 - 1)$,
\be
\dot \Psi_{\rm IR} =  -2\ell^2 \int \frac{d^dp}{(2\pi)^d}\,\delta(p^2 \ell^2 - 1) \left[ N  B(p,-p) + \int \frac{d^dq}{(2\pi)^d}\,\frac{d^dr}{(2\pi)^d}\, B(q,p) B(-p,r) \frac{\delta}{\delta B(q,r)}+ {\rm const.}\right] \Psi_{\rm IR} \,;
\ee
This works even if the full path integral is not gaussian, as the nongaussian terms are suppressed by powers of $d\ell$.

\subsection{Cutoffs on bilinears}

In the introduction, we have argued that the QFT cutoff is expected to act on the singlet fields, the bilinears
\be
\sigma(x,y) = \frac{1}{N} \phi^\dagger(x) \!\bracedot \!\phi(y) \,.
\ee
The idea of deriving the higher spin duality by writing the QFT in terms of these collective fields was initiated in Ref.~\cite{Das:2003vw}.
To write the path integral with the bilinears as independent fields one must include the Jacobian~\cite{Jevicki:1980zg,deMelloKoch:1996mj},
\be
{\cal D}\phi = {\cal D}\sigma \,J \,, 
\ee
where
\be
\ln J = N {\rm Tr} \ln \boldsymbol\sigma + O(N^0)\,. 
\ee
The bold $\boldsymbol\sigma$ is treated as a matrix.  In this paper we are interested in free fields or at most tree level in the bulk, so will drop the $O(N^0)$ terms henceforth.  Including the Jacobian, the unregulated Euclidean action is
\be
S = - N {\rm Tr} (\partial^2 \bsig + \ln \bsig) \,.
\ee
The linear term can be removed by a shift
\be
\bsig = \bsig_0 + \bchi \,, \quad \bsig_0 = - \partial^{-2} \bsig \,,
\ee
leaving
\be
S = - N {\rm Tr} \left(\ln \bsig_0 - \frac{1}{2} \bsig_0^{-1} \bchi \bsig_0^{-1} \bchi + \frac{1}{3} 
 \bsig_0^{-1} \bchi \bsig_0^{-1} \bchi \bsig_0^{-1} \bchi + \dots
\right) \,.
\label{collact}
\ee

Writing out the quadratic term expicitly,
\be
S_{(2)} = \frac{N}{2} \int \frac{d^dp}{(2\pi)^d}\,\frac{d^dq}{(2\pi)^d}\, p^2 q^2 \chi(p,q) \chi(-q,-p) \,.
\ee
We see that writing the action in terms of the bilinear field automatically introduces double-bilinears in the action.  If we now impose cutoffs on $p$ and $q$ separately, we will get an RG equation equivalent to the one in \S\ref{HRGVector}, but there is now the possibility of more general cutoffs.\footnote{Ref.~\cite{Zayas:2013qda} added multi-bilinears to the action, but with a single-bilinear cutoff.}

For example, we can impose a hard cutoff on the internal momentum $Q$ of the bilinear.  Defining the center of mass $R = (x_1+x_2)/2$ and the separation $r = x-y$, the conjugate momenta are $P = p+q$ and 
$
Q = (p-q)/2.
$
The cut-off path integral is then
\be
\Psi_{\rm IR}(\ell,B) = \int {\cal D}\chi\Bigl|_{|Q| \ell < 1} \, \exp \left\{-S_{(2)} + N {\rm Tr}(\bB \bchi) \right\} \,. \label{hsir}
\ee
The interaction terms can be reintroduced by a Legendre transformation as in~(\ref{legendre}).  Integrating out an infinitesimal shell as described at the end of \S\ref{HRGVector}, we obtain
\be
\dot \Psi_{\rm IR} =  -N \ell^2 \int \frac{d^dp}{(2\pi)^d}\,\frac{d^dq}{(2\pi)^d}\, \delta(Q^2 \ell^2 - 1) \left[ \frac{1}{p^2 q^2} B(p,q)B(-q,-p) + {\rm const.}\right] \Psi_{\rm IR} \,.
 \label{rgQ}
\ee
Unfortunately this is still not satisfactory.  First, it still does not have the desired structure $B^2 + \delta_B^2$ that leads to a second order equation.  The first step, analogous to Eq.~(\ref{rgtriv}), would give $\dot \Psi_{\rm IR} \sim \chi^2 \sim \delta_B^2$, but the remaining steps using the Schwinger-Dyson equation remove this.  Second, it is badly nonlocal, with poles at $p^2 =0$ and $q^2 = 0$.  

We have tried various strategies to modify the cutoff.  The most interesting is to cut the path integral off in the separation, $|r| > \ell$, as discussed in the introduction.  Writing the bilinears and sources now as functions of $P = p+q$ and $r$, the quadratic action is
\be
S_{(2)} = \frac{N}{2} \int \frac{d^dP}{(2\pi)^d}\,d^dr\, \chi(P,r) (P/2+i\partial_r)^2 (P/2-i\partial_r)^2 \chi(-P,r) \,. \label{covrg}
\ee
The presence of derivatives with respect to $y$ complicates the treatment of a hard cutoff.  In order for the path integral to be well-defined, we need to specify $\chi(P,r)$ and $\partial_{|r|} \chi(P,r)$ at the boundary $|r|=\ell$, and these values will flow as the cutoff is lowered.

A similar structure arises in a somewhat simpler way in the light-cone version of the bilinear field theory.  We therefore turn to this next.

\sect{Higher spins on the light cone}
\label{HSLC}
Ref.~\cite{Koch:2010cy} showed light-cone frame in combination with the collective field method allowed for a simple construction of  `precursors,'  bulk fields in terms of the boundary operators, at least to lowest order in $1/N$.  
In this section we develop the RG for the light-cone form of the theory.

\subsection{Bilocal fields on the light cone}

We work with coordinates
\be
ds^2 = 2dx^+ dx^- + d\vec x \cdot  d\vec x \equiv 2dx^+ dx^- + d x^i   d x^i  \,.
\ee
Ref.~\cite{Koch:2010cy} rewrites the vector theory in terms of an equal time bilocal field, which is defined by
\be
\sigma(x^+,x,y) \equiv \phi^{a\dagger}(x^+,x^-,\vec x) \phi^a(x^+, y^-, \vec y) \, .
\ee
We are working here with the $U(N)$ theory, and can truncate to the $O(N)$ theory by making $\phi^a$ real.

We will derive the collective field theory in Hamiltonian form.  The free action is
\be
S = -\int d^dx\left( 2 \partial_+ \phi^{a\dagger}   \partial_- \phi^a + \vec \partial \phi^{a\dagger} \cdot \vec\partial \phi^a \right)\,.
\ee
This gives the equal-time commutator (we suppress the equal time argument $x^+$)
\be
[ \phi^{a\dagger}(p^+, \vec x), \phi^b(q^+, \vec y)] = \frac{\pi}{p^+} \delta^{ab} \delta(p^+ + q^+) \delta^{d-2}(\vec x - \vec y) \,.
\ee
We will always take $p^+$ to be positive, and define
\be
\pi^a(p^+, \vec x) = 2ip^+ \phi^{a\dagger}(-p^+, \vec x) \,,\quad \pi^{a\dagger}(p^+, \vec x) = 2ip^+ \phi^{a}(-p^+,\vec x) \,.
\ee
Here $\phi^{a\dagger}(-p^+, \vec x) = (\phi^{a}(p^+, \vec x))^\dagger$.
The energy of a mode is 
\be
p^- = -\frac{\vec p \cdot \vec p}{2p^+} \, ,
\ee
and so the Hamiltonian is
\be
P^- = -i \int_0^\infty \frac{dp^+}{2\pi}\int \frac{ \transd \vec p}{(2\pi)^{d-2}} \,p^- \left[\pi^a(p^+, \vec p)  \phi^a(p^+, \vec p) + \pi^{a\dagger}(p^+, \vec p)  \phi^{a\dagger}(p^+, \vec p)\right] \,.
\ee

We now introduce the bilocal field
\be
\chi(p_1^+, \vec x_{ 1}, p_2^+, \vec x_{ 2}) = \frac{\sqrt{4p_1^+ p_2^+}}{N} \phi^{a\dagger}(p_1^+, \vec x_{ 1})  \phi^a( p_2^+, \vec x_{ 2}) \,,
\ee
and we define the canonical momentum
\be
\left[\pi_\chi(p_1^+, \vec x_{ 1}, p_2^+, \vec x_{ 2}), \chi(p_3^+, \vec x_{ 3}, p_4^+, \vec x_{ 4}) \right] = -i (2\pi)^2 \delta(p_1^+ - p_3^+) \delta^{d-2}(\vec x_{ 1} - \vec x_{ 3}) \delta(p_2^+ - p_4^+) \delta^{d-2}(\vec x_{ 2} - \vec x_{ 4})   \,.  \label{Pichi}
\ee
For compactness we will write this later as
\be
\left[\pi_{\chi 12}, \chi_{34} \right] = -i \delta_{13} \delta_{24} \,.
\ee
The Hamiltonian is
\be
P^- = -i \int \frac{dp_1^+ \transd \vec p_{ 1} dp_2^+ \transd \vec p_{ 2}}{(2\pi)^{2d-2}} \,(  p^-_1 + p^-_2) \pi_\chi(p_1, p_2)\chi(p_1, p_2) \,.
\label{lcham}
\ee
To be precise, this is correct acting on $U(N)$-invariant states constructed by acting on the vacuum with the bilinear creation operator $\pi_\chi$.  For baryonic states constructed with the $\epsilon$ tensor we would need another term, but we will not consider these.
The Hamiltonian path integral is then
\be
\int {\cal D}\chi\, {\cal D}\pi_\chi \exp  \int dx^+ \frac{dp_1^+ \transd \vec p_{ 1} dp_2^+ \transd \vec p_{ 2}}{(2\pi)^{2d-2}}\,
\pi_\chi (p_1,p_2)(i \partial_+ - p_1^- - p_2^-) \chi (p_1,p_2)  \,.
\ee

This looks like a free path integral, but there is a nonlinearity hidden in the reality condition on $\chi$ and $\pi_\chi$.  Consider the conjugate
\be
\chi(p_1^+, \vec x_{ 1}, p_2^+, \vec x_{ 2})^\dagger = \frac{\sqrt{4p_1^+ p_2^+}}{N}  \phi^a(-p_1^+, \vec x_{ 1})  \phi^{a\dagger}( -p_2^+, \vec x_{ 2}) \,.
\ee
Its commutator with $\chi$ is
\be
\bigl[\chi_{12}^\dagger, \chi_{34} \bigr] = -\frac{1}{ N} \delta_{13} \delta_{24} - \frac{\sqrt{4p_2^+ p_4^+}}{N^2} \delta_{13}
\phi^{a\dagger}( -p_2^+, \vec x_{ 2}) \phi^a( p_4^+, \vec x_{ 4})  - \frac{\sqrt{4p_1^+ p_3^+}}{N^2 } \delta_{24}
\phi^a( -p_1^+, \vec x_{ 1})  \phi^{a\dagger}( p_3^+, \vec x_{ 3})  \,.
\label{cdc}
\ee
The first term, leading in $1/N$, is from double commutators, and it is a multiple of the $\pi_\chi\chi$ commutator~(\ref{Pichi}).  At the level of free fields in the bulk this is all we need, so to this order the action is simply quadratic in $\chi, \chi^\dagger$.  However, we will want to discuss tree level bulk interactions in \S\ref{Conclusion}, so we will develop the single-commutator terms somewhat further.

The single-commutator terms involve a field of positive frequency contracted into a field of negative frequency, and so are not immediately expressed in terms of $\chi$ and $\chi^\dagger$.  To do so, note that the negative-frequency modes annihilate $\langle 0|$, and so
\bea
\langle 0 | \chi_{34} \phi^{a\dagger}( -p_1^+, \vec x_{ 1})  \phi^a( p_2^+, \vec x_{ 2}) &=& -\frac{1}{\sqrt{4p_1^+ p_2^+}} \delta_{14} \langle 0 | \chi_{32} \nonumber\\
&=& \frac{i}{\sqrt{4p_1^+ p_2^+}}\int \frac{dp^+_5 }{2\pi} \transd \vec{x}_5\langle 0 | \chi_{34} \pi_{\chi51}\chi_{52} \nonumber\\
&\equiv& \frac{i}{\sqrt{4p_1^+ p_2^+}} \langle 0 | \chi_{34} \pi_{\chi A1}\chi_{A2}
\,.
\eea
In the last line we have introduced a condensed notation for the integral.  This extends to arbitrary products of bilinear excitations on $\langle 0 |$, so we have
\be
 \phi^{a\dagger}( -p_1^+, \vec x_{ 1})  \phi^a( p_2^+, \vec x_{ 2})  = \frac{i}{\sqrt{4p_1^+ p_2^+}} \pi_{\chi A1}\chi_{A2} \,,\quad
 \phi^a( -p_1^+, \vec x_{ 1})  \phi^{a\dagger}( p_2^+, \vec x_{ 2})  = \frac{i}{\sqrt{4p_1^+ p_2^+}} \pi_{\chi 1A}\chi_{2A} \,.  \label{pmpc}
\ee

Combining the result~(\ref{pmpc}) with the commutator~(\ref{cdc}), one finds that
\be
\left[\chi^\dagger_{12}, \chi_{34} \right] = \left[ \left(\frac{-i}{N} \pi_{\chi 12} + \frac{1}{N^2} \pi_{\chi A2} \pi_{\chi 1B} \chi_{AB} \right), \chi_{34} \right]\,.
\ee
This implies the operator equation
\be
\chi^\dagger_{12} =  - \frac{i}{N} \pi_{\chi 12} + \frac{1}{N^2} \pi_{\chi A2} \pi_{\chi 1B} \chi_{AB}  \,. \label{cppcp}
\ee
Any additional term would have to commute with $\chi_{34}$ and so contain only positive frequencies, but all terms here have negative total frequency.  This form has a simple physical description.  Acting to the left, $\chi$ creates a bilocal two particle mode, while by definition $\pi_\chi$ destroys such a mode.  The operator $\chi^\dagger$ instead destroys two separate $\phi$'s.  If these two $\phi$'s are part of the same bilocal mode, it acts identically to $\pi_\chi$.  If these two $\phi$'s are part of different bilocal modes, it destroys both of them and creates a new bilocal mode out of the two remaining $\phi$'s.   

Eq.~(\ref{cppcp}) can be inverted to write $\pi_\chi$ in terms a $\chi$ and $\chi^\dagger$.  One finds
\bea
\pi_\chi 
&=& i N \left( \chi^\dagger  +  \!:\! \chi^\dagger \chi^{\rm T} \chi^\dagger \!: + 2  \!:\! \chi^\dagger \chi^{\rm T} \chi^\dagger \chi^{\rm T} \chi^\dagger \!: + \ldots  \right)
\,. \label{pinonlin}
\eea

Inserting this into the light-cone Hamiltonian~(\ref{lcham}), the interactions become explicit.  Note that only terms with even numbers of collective fields appear.  Correlators with odd number of collective fields are still nonzero, due to the identities~(\ref{pmpc}).  By contrast, the covariant action~(\ref{collact}) contains all powers of the field; it would be interesting to relate these two actions directly.  The quadratic part of the Hamiltonian is
\be
P_{(2)}^- = N \int \frac{dp_1^+ \transd \vec p_{ 1} dp_2^+ \transd \vec p_{ 2}}{(2\pi)^{2d-2}} \, (  p^-_1 + p^-_2) \chi^\dagger(p_1, p_2)\chi(p_1, p_2) \,.
\label{quad}
\ee

\subsection{Light-cone RG}
\label{HSRG}

We now consider the RG flow of the Hamiltonian path integral with the action~(\ref{quad}).  We wish to impose a hard cutoff on the separation in the bilinear.  Noting the effectively nonrelativistic light-cone kinematics, we define the center of mass variable and the separation 
\be
\vec x = (p_1^+ \vec x_1 + p_2^+ \vec x_2)/P^+ \,,\quad 
\vec r = \vec x_1 - \vec x_2 \,,
\label{cm}
\ee
 with conjugate momenta again $P$ and $Q$ respectively.  Then
\be
 p^-_1 + p^-_2 =  \frac{\vec p^{\, 2}_{ 1}}{2 p_1^+} + \frac{\vec p^{\, 2}_{\vec 2}}{2 p_{\, 2}^+} = \frac{\vec P^{2}}{2 P^+} + \frac{\vec Q^{\, 2}}{2 \mu} \,,
\ee
where the reduced mass is $\mu = p_1^+ p_2^+/P^+$.  The quadratic part of the action is
\bea
S_{(2)} = N \int dx^+ \frac{dp_1^+ dp_2^+ \transd \vec x\, \transd \vec r}{(2\pi)^{2}}&&\left[
\chi(p_1^+,p_2^+, \vec x, \vec r)^\dagger\biggl(i\partial_+ + \frac{\partial_{\vec x}^{ 2}}{2P^+} \biggr) \chi (p^+_1,p^+_2, \vec x, \vec r) \right.
\nonumber\\
&&\left. -\frac{1}{2\mu} \partial_{\vec r}\chi(p^+_1,p^+_2, \vec x, \vec r)^\dagger \cdot \partial_{\vec r} \chi (p^+_1,p^+_2, \vec x, \vec r) \right]\,.
\eea

We want to define the path integral $\Psi_{\rm IR}(\ell,B)$ with a hard cutoff, restricting the fields to $|\vec r| > \ell$.  Actually, we will see that the cutoff needs to be slightly different from this, so we will put ?'s in equations that will need to change.  As discussed in \S\ref{HRGVector}, the $\vec r$ derivatives in the action require boundary conditions to give a well defined path integral.  We must specify either $ \chi(p^+_1,p^+_2, \vec x, \vec r)$ or $\vec r \cdot \partial_{\vec r} \chi (p^+_1,p^+_2, \vec x, \vec r)$ at $|\vec r| > \ell$.  Moreover, this boundary value will run as $\ell$ is varied, so we must include it as an argument of $\Psi_{\rm IR}$.  In fact, varying the boundary value of $\chi$ or its normal derivative is equivalent to adding a delta-function source at $|\vec r| > \ell + \epsilon$.  We will therefore use the boundary value in lieu of the general source $B$, identifying it with the bulk field:
\be
\Psi_{\rm IR}(\ell,\tilde\chi) \stackrel{?}= \int {\cal D}\chi^\dagger {\cal D}\chi\bigl|_{\chi(|\vec r| = \ell) = \tilde \chi} \, e^{iS_{(2)}} \,.
\label{lcpsi}
\ee

When $\ell$ is varied, $\Psi_{\rm IR}$ evolves under the radial Hamiltonian $H$, not to be confused with the light-cone Hamiltonian $P_-$:
\be
\partial_\ell \Psi_{\rm IR}(\ell,\tilde\chi) = i H \Psi_{\rm IR}(\ell,\tilde\chi) \,.
\ee
This Lorentzian equation includes a factor of $i$, in contrast to the Euclidean~(\ref{ham}).
The radial Hamiltonian is obtained in the usual way from the action.  To quadratic order it is
\bea
&& \quad H_{(2)} \stackrel{?}{=} \int dx^+ \frac{dp_1^+ dp_2^+ \transd \vec x_{ }\, \transd \vec r}{(2\pi)^{d}}\left[
N \chi(p_1^+,p_2^+, \vec x, \vec r)^\dagger\biggl(i\partial_+ + \frac{\partial_{\vec x}^{ 2}}{2P^+} \biggr) \chi (p^+_1,p^+_2, \vec x, \vec r)   \right.
\nonumber\\
&& + \left.\frac{N}{2\mu} \partial_{\vec r}\chi(p^+_1,p^+_2, \vec x, \vec r)^\dagger \cdot {\rm P}^\bot\cdot\partial_{\vec r} \chi (p^+_1,p^+_2, \vec x, \vec r)  -\frac{2\mu}{N} \Pi(p^+_1,p^+_2, \vec x, \vec r)^\dagger \Pi (p^+_1,p^+_2, \vec x, \vec r) \right]\,. \qquad\label{lcradham}
\eea
Here ${\rm P}^\bot$  projects onto the directions perpendicular to $\vec r$.  As reviewed in \S\ref{HRGReview}, the RG equation for $\tilde \chi$ is simply the Hamiltonian equation of motion~(\ref{heis}), or its Lagrangian equivalent.  Here that would be
\be
\left(i\partial_+ +\frac{ \partial_{\vec x}\cdot \partial_{\vec x}}{2P^+}+  \frac{\partial_{\vec r}\cdot \partial_{\vec r}}{2\mu} \right) \chi (p^+_1,p^+_2, \vec x, \vec r)\ = 0,.
\ee

However, we are not quite done.  We expect the coefficients of the Poincar\'e and radial second derivatives to be equal in the AdS background.  In order to accomplish this, we need to define the radial coordinate as 
\be
z =  |\vec r| \sqrt{\mu/P^+} = |\vec r|  \sqrt{p_1^+ p_2^+}/P^+ \,.
\ee
As we will review further in \S4, this identification has already been made in the precursor language~\cite{Brodsky:2006uqa, Koch:2010cy, Jevicki:2011ss}.

We can now correct the tentative equations above.  In the path integral~(\ref{lcpsi}), the uniform cutoff in $z$ means that the cutoff in field space is at $|\vec r| = \ell P^+/\sqrt{p_1^+ p_2^+}$.  In the Hamiltonian~(\ref{lcradham}), the coefficient of the $\Pi^2$ term contains an additional factor of $P^{+2}/p_1^+ p_2^+$ after converting to the correct radial time.  Defining the vector
\be
\vec z = \vec r \frac{\sqrt{p_1^+ p_2^+}}{P^+}  \,, \label{vecz}
\ee
the equation of motion is
\be
\left(i\partial_+ +\frac{ \partial_{\vec x}\cdot \partial_{\vec x} + \partial_{\vec z}\cdot \partial_{\vec z} }{2P^+} \right) \chi (p^+_1,p^+_2, \vec x, \vec r)\ = 0 \,. \label{rgeom}
\ee

\subsection{Higher spins on the light cone}

We now wish to make contact with the light-cone description of the higher spin bulk fields.  Since we are studying at this point free bulk fields, we can work with the Fronsdal description~\cite{Fronsdal:1978rb} rather than the full Vasiliev theory.  The light-cone gauge has been extensively developed by Metsaev~\cite{Metsaev:1999ui}.  We review here the language and some of the main results of that work.

The spin-$S$ Fronsdal higher spin field $h_{\mu_1 \mu_2 \dots \mu_S}$ is symmetric with a double tracelessness condition
\be
\tensor{h}{_\nu}{^\nu}{_\rho}{^\rho}{_{\mu_5\dots \mu_S}} = 0 \, .
\ee
For the $U(N)$ theory we have all integer spins, while for the $O(N)$ theory we truncate to even spins.  These have a gauge invariance
\be
h^\prime_{\mu_1 \dots \mu_S} = h_{\mu_1 \dots \mu_S} + \nabla_{(\mu_1} \Lambda_{\mu_2 \dots \mu_S)} \, ,
\ee
for a symmetric traceless gauge parameter $\Lambda_{\mu_1 \dots \mu_{S-1}}$ and the AdS covariant derivative $\nabla_\mu$.  

The Fronsdal fields can be collected into a single field via the generating functional 
\be
\Phi = \sum_{s=0}^\infty h_{A_1 \dots A_S} \alpha^{A_1} \dots \alpha^{A_S} |0\rangle \, .
\label{HSGenFunc}
\ee
where $\alpha^{A}$ is an auxiliary variable the acts as a creation operator.  The lowering operator is  $\bar{\alpha}^A$, so that
\be
[\bar{\alpha}^A, \alpha^B] = \eta^{AB} \, .
\ee
Indices $A, B$ run over all frame field dimensions, and further decompose
\be
A \to (+, -, I) \to (+, -, z, i) \,.
\ee
As usual, the AdS vierbein $e^A_\mu$ may be used to convert frame field indices to spacetime indices.  For this entire paper, we will be working in the Poincar\'{e} patch of AdS, so
\be
e^A_\mu = \frac{1}{z} \delta^A_\mu \, .
\ee

In this language the double traceless condition is
\be
\bar{\alpha}^2 \bar{\alpha}^2 \Phi = 0 \, .
\ee
A gauge transformation is
\be
\Phi^\prime  = \Phi + \alpha^A D_A \Lambda \, .
\ee
for $\bar{\alpha}^2 \Lambda = 0$.   Here $D_A$ is the AdS covariant derivative in frame field indices, which may be written
\be
D_A \equiv \hat{\partial}_A + \frac{1}{2} \omega^{BC}_A M^{BC} \, .
\ee
where the Lorentz spin operator $M^{BC}$ may be expressed in terms of $\alpha$ and $\bar{\alpha}$ as
\be
M^{BC} = \alpha^B \bar{\alpha}^C - \alpha^C \bar{\alpha}^B \, .  \label{spin}
\ee
Also, $\hat{\partial}_A \equiv e_A^\mu \partial_\mu$ and $\omega^{BC}_A$ is the frame field spin connection for Poincar\'{e} AdS.  The equation of motion for the higher spin fields is then
\be
\label{HSEoMNoGauge}
\left ( D_A^2 + \omega^{AAB} D_B - S^2 + (5-d)S + 2d - 4 - \alpha D \bar{\alpha}D + \frac{1}{2} (\alpha D)^2 \bar{\alpha}^2 - \alpha^2 \bar{\alpha}^2\right ) \Phi = 0 \, .
\ee
The spin $S$ may be written as $S= \alpha^A \bar{\alpha}_A$, the above uses the abbreviation $\alpha D \equiv \alpha^A D_A$, and the dimension $d$ is that of the boundary theory; this differs from \cite{Metsaev:1999ui}, in which $d$ is the bulk dimension.  In light-cone gauge, it can then be shown that $\Phi$ is single-traceless, so $\bar{\alpha}^2 \Phi = 0$.  Then, the equations of motion may be rewritten as
\be
\left ( z^2 \partial^A \partial_A + \frac{1}{2} M_{ij}^2 - \frac{(d-3)(d-5)}{4} \right ) \frac{\Phi}{z^{(d-1)/2}}  = 0 \, ,
\label{lceom}
\ee
where again $i,j$ run over non-$z$, non-light-cone coordinates.  In particular, for $d=3$, this just
\be
z^2 \partial^A \partial_A \left ( \frac{\Phi}{z} \right ) = 0 \, .
\label{3DHSLCEoM}
\ee

To compare with the RG equation~(\ref{rgeom}), note that
\be
\partial_{\vec z}\cdot \partial_{\vec z} = \partial_z^2 + \frac{d-3}{z} \partial_z  + \frac{1}{2z^2} M_z^{ij} M_z^{ij}  \,.
\ee
Here the $SO(d-2)$ spin $M_z^{ij} = z^i \partial_{z^j} - z^j \partial_{z^i}$ is antihermitean, consistent with the bulk convention~(\ref{spin}).  Using also $\partial^A \partial_A = 2 i P^+ \partial_+  + \partial_{\vec x} \cdot \partial_{\vec x} + \partial_z^2$, one finds that the RG equation for $z^{(d-3)/2} \chi$ is the same as the bulk equation~(\ref{lceom}) for ${\Phi}/{z^{(d-1)/2}} $.  In other words, we have found the correct second order equation from the RG provided that we identify
\be
\chi = z^{2-d} \Phi \,.  \label{chiphi}
\ee

\subsection{Matching RG and higher spin fields}
\label{MatchRGHS}

The matching of equations of motion is satisfying but not complete.  We see that fields with the same $SO(d-2)$ spin match, but this is not a complete characterization.

The RG field is $\chi(p_1^+, p_2^+, \vec x, \vec z)$.  The light-cone bulk field is $\Phi(P^+, \vec x, \alpha^z, \vec \alpha)$.\footnote{All vectors are in the $d-2$ transverse dimensions.}  To facilitate comparison, let us introduce a parameter~$\theta$~\cite{Koch:2010cy} through 
\be
p_1^+ = P^+ \cos^2 (\theta/2)\,,\quad p_2^+ = P^+ \sin^2 (\theta/2)\,, \label{deftheta}
\ee
in terms of which
\be
\vec z= \frac{\vec x_1 - \vec x_2}{2}  \sin\theta  \,. \label{defvecz}
\ee
  Also, let us consider fields of $SO(d-2)$ spin $J$:
\bea
\chi(p_1^+, p_2^+, \vec x, \vec z) &=& \chi_J(P^+, \vec x, z, \theta) z^{-J} T^J(\vec z)  \,,  \nonumber\\
\Phi(P^+, \vec x, z, \alpha^z, \vec \alpha) &=& \Phi_J(P^+, \vec x, z, \alpha^z, \vec \alpha \cdot \vec\alpha) T^{J}(\vec \alpha) \,.
\eea
The traceless homogeneous order-$J$ polynomials $T^{J}$ are the same on both sides to match the $SO(d-2)$ spins.  There remain one extra variable $\theta$ in $\chi$ and two extra variables $\alpha^z, \vec \alpha \cdot \vec\alpha$ in $\Phi$.  One of the latter is removed by the tracelessness condition 
\be
\bar\alpha^I \bar \alpha^I \Phi |0\rangle = 0\,. \label{tless}
\ee
To find the correspondence between the remaining variables, we match transformations under the boost $J^{-i}$.  This has already been carried out in the precursor context for $d=3$~\cite{Koch:2010cy}, and partially for higher $d$~\cite{Jevicki:2011ss}; here we complete the latter exercise.

First, we impose the tracelessness condition.  Take $\Phi_J$ to be of the form
\be
\Phi_J(P^+, \vec x, z, \alpha^z, \vec \alpha \cdot \vec\alpha) = \Phi_{S,J}(P^+, \vec x, z) f_{S,J}(\alpha^z, \vec \alpha \cdot \vec\alpha) \label{bulkDecompose}
\ee
with $f_{S,J}(\alpha^z, \vec \alpha \cdot \vec\alpha)$ of order $S-J$ in $\alpha^I$.  The tracelessness condition becomes
\be
\left( \partial_u^2 + 4v \partial_v^2 + (2d + 4J - 4) \partial_v \right) f_{S,J}(u, v) = 0 \,,
\ee
where we abbreviate $\alpha^z = u$ and $ \vec \alpha \cdot \vec\alpha = v$.
The solution is the polynomial
\be
f_{S,J}(u, v) = \sum_{m=0}^{\lfloor (S-J)/2 \rfloor}
\frac{u^{S-J-2m} v^m}{(-4)^m m!} \frac{(S-J)!}{(S-J-2m )!} \frac{(J - 2 + d/2)!}{(J - 2 + m + d/2)!} \,.
\ee
Note that $S$ and $J$ respectively are the $O(d-1)$ and $O(d-2)$ spins,
\be
M^{IJ} M^{IJ} = -2 S (S + d - 3) \,,\quad M^{ij} M^{ij} = - 2J (J + d - 4) \,,
\ee
and that $S \geq J \geq 0$.
Note also that $d=3$ is special, in that $M_{ij} = 0$; we will give this separate treatment at the end.

We expect a $\Phi \to \chi$ mapping
\be
\Phi = \sum_{S,J} \Phi_{S,J}(P^+, \vec x, z) f_{S,J}(u, v) T^{J}(\vec \alpha) 
\ \Longleftrightarrow\ \chi = z^{2-d} \sum_{S,J} \Phi_{S,J}(P^+, \vec x, z) g_{S,J}(\theta) z^{-J}  T^{J}(\vec z)  \,,
\label{ftog}
\ee
including the factor~(\ref{chiphi}).
The functions $g_{S,J}(\theta)$ are determined by the requirement that $J^{-z}$ act in the same way on both sides; that is, it commutes with the mapping $\Phi \to \chi$.  To make things simpler, Ref.~\cite{Jevicki:2011ss} has already separated $J^{-i}$ into an orbital part, which maps easily, and a spin part that depends on $M^{iz}$.  The latter acts on $\Phi$ as
$ M^{iz}|_\Phi =  \alpha^i \bar{\alpha}^z - \alpha^z \bar{\alpha}^i$, and on $\chi$ as~\cite{Jevicki:2011ss}
\bea
M^{iz}|_\chi &=&i \frac{x_1^i-x_2^i}{\sqrt{(x_1^j-x_2^j)^2}} \left[\sqrt{p_1^+ p_2^+}(x_1^--x_2^-)+\frac{\bigl((p_1^+)^2 p_2^j+(p_2^+)^2 p_1^j \bigr)(x_1^j-x_2^j)}{P^+\sqrt{p_1^+ p_2^+}}\right] \cr
&&\qquad\qquad +\, \frac{1}{2}\frac{p_1^+-p_2^+}{P^+}\sqrt{(x_1^j-x_2^j)^2}\left(p_1^i\sqrt{p_2^+ \over p_1^+}-p_2^i\sqrt{p_1^+ \over p_2^+}\right) \,.
\eea
(The operator ordering in the last term on the first line has been corrected.)
In terms of the variables $\theta$, $\vec z$ this simplifies   to
\be
M^{iz}|_\chi =  \hat z^i \partial_\theta + \cot \theta \,{\rm P}^{ij} z \partial_{z^j} \,,
\ee
where P again projects into the space orthogonal to $\vec z$.

Acting on $\Phi$, 
\bea
M^{iz}|_\Phi f_{S,J}(u, v) T^{J}(\vec \alpha) |0\rangle&=& \left[ ( \partial_u - 2 u  \partial_v )  f_{S,J}(u, v) \right]\alpha^i  T^{J}(\vec \alpha) |0\rangle - u f_{S,J}(u, v) \partial_{\alpha^i} T^{J}(\vec \alpha) |0\rangle \nonumber\\
&=&  \left[ ( \partial_u - 2 u  \partial_v )  f_{S,J}(u, v) \right] [\alpha^i  T^{J}(\vec \alpha)]_{\rm traceless} |0\rangle \nonumber\\
&&\quad + \left[ \frac{v ( \partial_u - 2 u  \partial_v ) }{d+2J - 4} - u\right]  f_{S,J}(u, v)  \partial_{\alpha^i} T^{J}(\vec \alpha) |0\rangle
\nonumber\\
&=&  \frac{(J+S+d-3)(S-J)}{d + 2 J-2}  f_{S,J+1}(u, v) [\alpha^i  T^{J}(\vec \alpha)]_{\rm traceless} |0\rangle \nonumber\\
&&\quad -  f_{S,J-1}(u, v)  \partial_{\alpha^i} T^{J}(\vec \alpha) |0\rangle
\eea
Each $f_{S,J\pm 1}$ multiplies a traceless polynomial of order $J \pm 1$, as required by the traceless condition~(\ref{tless}).  The mapping~(\ref{ftog}) takes this to
\be
 \frac{(J+S+d-3)(S-J)}{d + 2 J-2}  g_{S,J+1}(\theta) z^{-J-1}[ z^i  T^{J}(\vec z)]_{\rm traceless}  -  g_{S,J-1}(\theta) z^{-J+1} \partial_{i} T^{J}(\vec z) \,.
 \ee

 This must be equal to 
 \bea
 M^{iz}|_\chi g_{S,J}(\theta)z^{-J} T^{J}(\vec z) &=& (\partial_\theta - J \cot \theta) g_{S,J}(\theta) z^{-J-1} \vec z^{\,i} T^J(\vec z)
 + \cot\theta \, g_{S,J}(\theta)z^{-J+1} \partial_i T^J(\vec z)
 \nonumber\\ 
 &=&  (\partial_\theta - J \cot \theta) g_{S,J}(\theta) z^{-J-1} [\vec z^{\,i} T^J(\vec z)]_{\rm traceless} \nonumber\\
 &&\quad
 + \left( \frac{\partial_\theta - J \cot \theta}{d+2J - 4}  + \cot\theta\right) g_{S,J}(\theta)  z^{-J+1} 
 \partial_i T^J(\vec z)
  \,.
 \eea
 Then
\bea
 (\partial_\theta - J \cot \theta) g_{S,J}(\theta) &=&  \frac{(J+S+d-3)(S-J)}{2 J-2+d}  g_{S,J+1}(\theta) \,,
 \quad J \geq 0\,, \label{raise}\\
 \partial_\theta g_{S,J} + (d + J - 4) \cot \theta\, g_{S,J}  &=& - (d+2J-4) g_{S,J-1}(\theta) \,,
 \quad J \geq 1 \label{lower}\,.
 \eea
 Eq.~(\ref{raise}) for $J=S$ gives 
 \be
 g_{S,S}(\theta) = c_S \sin^S \theta \,,
 \label{JequalsS}
 \ee
 with a normalization not fixed by symmetry.  
 We may combine these equations to get the second order differential equation for $g_{S,J}$
 \be
 \partial^2_\theta g_{S,J} + (d-3) \cot \theta \partial_\theta g_{S,J} + S(S+d-3) g_{S,J} - \frac{J(J+d-4)}{\sin^2 \theta} g_{S,J} = 0 \, .
 \ee
 This differential equation has the solution
 \be
 g_{S,J}(\theta) = c_{S,J} (\sin \theta)^{-(d/2-2)} P^{J-2+d/2}_{S-2+d/2}(\cos \theta) \, ,
 \ee
 where $P^m_l(x)$ is the associated Legendre function.  The second solution, related to the independent associated Legendre function $Q^m_l(x)$, has been dropped since it is not compatible with the $J=S$ solution in Eq.~(\ref{JequalsS}).  We can then use Eq.~(\ref{lower}) to relate the constants $c_{S,J}$ for equal $S$ but different $J$ to each other, giving the complete result
\be
g_{S,J}(\theta) = c_S (-1)^{-(S-2+d/2)} \frac{(S-J)! (2J-4 +d)!!}{(S+J+d-4)!} \left ( \sin \theta \right )^{-(d/2-2)} P^{J-2+d/2}_{S-2+d/2} (\cos \theta) \, . \label{gSJ}
\ee
This is normalized so that the result for $J=S$ is identical to Eq.~(\ref{JequalsS}).

This determines the full mapping~(\ref{ftog}) between bulk and RG fields, up to an $S$-dependent normalization factor that depends on conventions.  Appendix \ref{TracelessCurrents} calculates the boundary currents and provides a check that this mapping gives the correct results.

\subsubsection{$d=3$}

The case $d=3$ needs special attention.  The transverse rotation group is $O(1) = Z_2$, corresponding to reflections on $\vec z$ and $\vec x$, both of which are one-dimensional vectors.  The $Z_2$ spin $J$ takes only the values 0 and 1, corresponding to the spin wavefunctions 
\be
T^0 = 1 \,,\quad T^1 = \hat z = {\rm sgn}(x_1 - x_2) \,,
\ee
respectively.  The bulk wavefunctions simplify to
\be
f_{S,0}(u, v) = {\rm Re}\left( (\alpha^z + i \alpha^x)^S \right) \,,\quad f_{S,1}(u, v) \alpha_x = S^{-1}  {\rm Im}\left( (\alpha^z + i \alpha^x)^S\right) \,.
\ee
Since traceless $T^J$ vanish for $J > 1$ the only nonvacuous conditions on $g_{S,J}$ are (\ref{raise}) for $J=0$ and (\ref{lower}) for $J=1$:
\be
\partial_\theta g_{S,0}(\theta) = S^2 g_{S,1}(\theta) \,, \quad \partial_\theta g_{S,1}(\theta) = - g_{S,0}(\theta) \,. 
\ee
The $g_{S,J}$ are then linear combinations of $\cos S\theta$ and $\sin S\theta$.  

The precise dictionary can be determined by the behavior under the interchange of the two fields in the bilinear.  In the $O(N)$ case, $\phi^{a}(x_1) \phi^a(x_2)$ is even under the exchange $x^\mu_1 \leftrightarrow x^\mu_2$, and only even spins $S$ appear.  In the $U(N)$ case, even spins are even under  exchange and odd spins are odd under exchange.  The exchange corresponds to 
\be
\theta \to \pi-\theta\,\quad \vec z \to - \vec z \,,
\ee
under which
\be
\cos S\theta \to (-1)^S\cos S\theta \,,\quad \sin S \theta\to (-1)^{S+1} \sin S\theta \,,\quad T^J \to (-1)^J  T^J \,.
\ee
It follows that $T^0 \cos S\theta$ and $T^1 \sin S \theta$ have the correct exchange behavior $(-1)^S$, and so
\be
g_{S,0} = c_S \cos S\theta \,,\quad g_{S,1} = S^{-1} c_S \sin S\theta \,.
\ee
Forming linear combinations,
\bea
(\alpha^z \pm i \alpha^x)^S \to c_{S} \left(\cos S\theta \pm i \hat z \sin S\theta\right) = c_{S} e^{\pm iS\hat z  \theta}
 \,.
\eea
Expanding in a complete basis,
\be
|\Phi(P^+, \vec x, z) \rangle = \sum_{S = -\infty}^\infty \Phi_S(P^+, \vec x, z) (\alpha^z + i \,{\rm sgn}(S) \alpha^x)^{|S|}  | 0 \rangle
\label{phid3}
\ee
corresponds to 
\be
\chi(p_1^+, p_2^+, \vec x_1, \vec x_2) = \frac{1}{z} \sum_{S = -\infty}^\infty c_{|S|} \Phi_S(P^+, \vec x, z) e^{ iS \hat z  \theta}  \,.\label{chid3}
\ee
There is a reality condition $\Phi_S^\dagger = \Phi_{-S}$.
We can express $\Phi$ directly in terms of $\chi$ by solving for the Fourier coefficients,
\bea
\Phi_S(P^+, \vec x, z) &= &\frac{z}{c_{|S|}} \int_0^{2\pi} \frac{d\theta}{2\pi} e^{- iS \hat z  \theta} \chi(p_1^+, p_2^+, \vec x_1, \vec x_2)   
\,.
\eea
Here, collecting earlier definitions, the arguments of $\chi$ in the integral are determined in terms of $(P^+, \vec x, z, \theta)$ by\footnote{The natural range of $\theta$ is from 0 to $\pi$, but these definitions give a natural extension to 0 to $2\pi$.}
\bea
p_1^+ = P^+ \cos^2 \frac{\theta}{2} \,,\qquad\qquad\qquad\qquad &&p_2^+ = P^+ \sin^2 \frac{\theta}{2} \,,
\nonumber\\
\vec x_1 = \vec x + \vec z \sqrt{\frac{p_2^+}{p_1^+}} =  \vec x + \vec z \tan\frac{\theta}{2} \,,\ \quad &&\vec x_2 = \vec x - \vec z \sqrt{\frac{p_1^+}{p_2^+}} =  \vec x - \vec z \cot\frac{\theta}{2}\,.
\eea
We can also write this as 
\bea
\Phi_S(P^+, \vec x, z) + \Phi_{-S}(P^+, \vec x, z)  &=& 
\frac{2z}{c_{S}} \int_0^{2\pi} \frac{d\theta}{2\pi}\, T_S \!\left( \frac{p_1^+ - p_2^+}{P^+} \right)\! \chi(p_1^+, p_2^+, \vec x_1, \vec x_2)  \,,  \nonumber\\
\Phi_S(P^+, \vec x, z) - \Phi_{-S}(P^+, \vec x, z)  &=& 
\frac{2 i \vec z}{c_{S}} \int_0^{2\pi} \frac{d\theta}{2\pi}\, \sin\theta\, U_{S-1} \!\left( \frac{p_1^+ - p_2^+}{P^+} \right)\! \chi(p_1^+, p_2^+, \vec x_1, \vec x_2) \,, \nonumber\\
\label{cheby}
\eea
where $T$ and $U$ are Chebyshev polynomials of the first and second kind.

Some particular components~(\ref{HSGenFunc}) of interest are
\bea
h_{x(S)}(P^+, \vec x, z) &=& i^S \Phi_S(P^+, \vec x, z) + i^{-S} \Phi_{-S}(P^+, \vec x, z) \,,
\nonumber\\
h_{zx(S-1)}(P^+, \vec x, z) &=& i^{S-1} \Phi_S(P^+, \vec x, z) + i^{-S+1} \Phi_{-S}(P^+, \vec x, z) \,, \label{hxz}
\label{phitoh}
\eea
where $(n)$ denotes an $n$-times repeated index.
Finally, we note again that for $d=3$ the equation of motion~(\ref{lceom}) conveniently simplifies to 
\be
\partial^A \partial_A  \left (\frac{\Phi}{z} \right )  = 0 \, .
\label{d3eom}
\ee

\sect{Higher spin precursors}
\label{HSP}
Our treatment in \S\ref{HSLC} is closely motivated by the work of Ref.~\cite{Koch:2010cy, Jevicki:2011ss}.  In particular our $d=3$ relation~(\ref{phid3}, \ref{chid3}) is essentially Eq.~(81) of~\cite{Koch:2010cy}.\footnote{We have tried to be slightly more precise about the dependence on $\hat z = {\rm sgn}(x_1 - x_2)$, and in using the exchange symmetry to distinguish $\alpha^z + i \alpha^x$ from $\alpha^x - i\alpha^z$.}  However, the context for these two results is slightly different.

The relation in~\cite{Koch:2010cy} defines a {\it precursor}, an operator in the CFT Hilbert space that is equal to a local bulk operator~\cite{Polchinski:1999yd}.  These are defined in the UV CFT without cutoff, whereas we are looking at operators in the theory with a Wilsonian cutoff. Clearly these are closely related in general, but here in the free CFT the constructions are identical, at least to linear order.

The precursors are usually described in terms of a pull-back, push-forward construction~\cite{Freivogel:2004rd}.  One first uses the bulk dynamics to express the bulk operators in terms of their boundary limits, and then the GKPW dictionary~\cite{GKP,W} to relate these limits to local CFT operators integrated against smearing functions that are nonlocal in time~\cite{BDHM,Balasubramanian:1998de,Bena99,Hamilton:2006az}.  The CFT evolution is used to express these in terms of operators on a single time slice, which are necessarily nonlocal in space.  In strongly coupled CFT's the last step cannot be carried out explicitly, but in the free CFT here it can, expressing the bulk fields in terms of CFT bilocals at a single time.  Further, the whole construction is determined by symmetry~\cite{Koch:2010cy}, up to the normalization factors that we have denoted $c_S$.

In the remainder of this section, we will elaborate certain aspects of the precursor point of view.

\subsection{Light-cone GKPW dictionary}
\label{LCDict}

As we have described above, the GKPW dictionary, relating boundary values of bulk fields to CFT operators, is one step in the standard precursor construction.   Since this dictionary has been derived directly from symmetry, we can go the other way and derive the light-cone form of the GKPW dictionary.  We will focus primarily on the case $d=3$ form simplicity.

Consider the $z\to 0$ behavior of the precursor construction (\ref{chid3}).  The LHS has a Taylor expansion in $\vec z$, and therefore so must the RHS.  In particular, the leading term will be of order $z^0$ and independent of $\hat z$, and the second term will be of order in $z$ and linear in $\hat z$.  These imply that 
\be
\Phi_S + \Phi_{-S} = O(z) \,,\quad \Phi_S - \Phi_{-S} = O(z^2) \,.  \label{lcexp}
\ee
As a check, these boundary behaviors are consistent with the nonnormalizable and normalizable modes of the field equation~(\ref{d3eom}).  

Comparing with the component expressions~(\ref{hxz}), we see that components with even numbers of $x$ indices are $O(z)$, and components with odd numbers of $x$ indices are $O(z^2)$.  In particular,
\bea
\lim_{z \to 0} \frac{1}{z} h_{x(2l)} (P^+, \vec x, z) &=&  \frac{4 (-1)^l}{c_{2l}} \int_0^{P^+} \frac{dp_1^+}{2\pi} \, T_{2l} \!\left( \frac{p_1^+ - p_2^+}{P^+} \right)\! \phi(p_1^+, \vec x ) \!\cdot\! \phi(p_2^+, \vec x )  \,, \nonumber\\
\lim_{z \to 0} \frac{1}{z^2} h_{x(2l+1)} (P^+, \vec x, z) &= & \frac{4 (-1)^{l+1}}{c_{2l+1}} \int_0^{P^+} \frac{dp_1^+}{2\pi}\, U_{2l} \!\left( \frac{p_1^+ - p_2^+}{P^+} \right)\! \frac{p_2^+ \partial_{\vec x_1} - p_1^+ \partial_{\vec x_2}}{P^+}
\phi(p_1^+, \vec x ) \!\cdot\! \phi(p_2^+, \vec x ) \,. \nonumber\\
\eea

From covariant versions of holography, one might have expected a local operator on the RHS, but there are two problems.  First, the integrand is polynomial in $p_+$ and $p_-$, but contains up to $S$ inverse powers of $P^+$.\footnote{We also have some residual uncertainty about the measure for the $p_1^+$ integral, which is related to operator ordering in the symmetry generators, but we have assumed that it works out to make the gauge-invariant $S=0$ terms local.}
 Second, the range of integration of $p_1^+$ is cut off.  In fact, the range can be extended to the full real line.  For $p_1^+ < 0$ or $p_1^+ > P^+$, one of $p_{1,2}^+$ is negative and so the operator annihilates the vacuum.  Thus it vanishes at the free bulk level that we are considering.  With this, we can write in position space
\bea
\lim_{z \to 0} \frac{1}{z} h_{x(2l)} (x^-, \vec x, z) &=&  \frac{4 (-1)^l}{c_{2l}}  T_{2l} \!\left( \frac{\partial_{x_1^-} - \partial_{x_2^-}}{\partial_{x_1^-} + \partial_{x_2^-}} \right)\! \phi(x^-, \vec x ) \!\cdot\! \phi(x^-, \vec x ) \,, \nonumber\\
\lim_{z \to 0} \frac{1}{z^2} h_{x(2l+1)} (x^-, \vec x, z) &= & \frac{4 (-1)^{l+1}}{c_{2l+1}} U_{2l} \!\left( \frac{\partial_{x_1^-} - \partial_{x_2^-}}{\partial_{x_1^-} + \partial_{x_2^-}}  \right)\! \frac{\partial_{x_2^-} \partial_{\vec x_1} - \partial_{x_1^-} \partial_{\vec x_2}}{i(\partial_{x_1^-} + \partial_{x_2^-})}
\phi(x^-, \vec x ) \!\cdot\! \phi(x^-, \vec x )\,. \nonumber\\ \label{gkpwlc}
\eea
Any field component with an even number of $z$ indices may be found easily from Eq.~(\ref{gkpwlc}) using the tracelessness of the bulk field.  To state the complete mapping we need also to write down the result for the field component with a single $z$ index.  This can be determined from Eq.~(\ref{cheby}) and Eq.~(\ref{phitoh}), and is
\bea
\lim_{z\to0} \frac{1}{z^2} h_{zx(2l - 1)}(P^+, \vec x, z) &=& \frac{4(-1)^{l}}{c_{2l}} U_{2l-1} \!\left( \frac{\partial_{x_1^-} - \partial_{x_2^-}}{\partial_{x_1^-} + \partial_{x_2^-}}  \right)\! \frac{\partial_{x_2^-} \partial_{\vec x_1} - \partial_{x_1^-} \partial_{\vec x_2}}{i(\partial_{x_1^-} + \partial_{x_2^-})}
\phi(x^-, \vec x ) \!\cdot\! \phi(x^-, \vec x ) \, , \nonumber  \\
\lim_{z\to 0} \frac{1}{z} h_{zx(2l)}(P^+, \vec x, z) &=& \frac{4 (-1)^{l}}{c_{2l+1}} T_{2l+1} \!\left( \frac{\partial_{x_1^-} - \partial_{x_2^-}}{\partial_{x_1^-} + \partial_{x_2^-}} \right)\! \phi(x^-, \vec x ) \!\cdot\! \phi(x^-, \vec x )\, . \label{gkpwlc2}
\eea
This is the light-cone form of the GKPW dictionary.  The RHS is $(\partial_{x_1^-} + \partial_{x_2^-})^{-S}$ acting on a local operator.   We will understand this in \S\ref{HSPdeDonder}.  For higher dimensions, a similar nonlocality may be found by unpacking the Legendre polynomial in Eq.~(\ref{gSJ}). 

\subsection{Relation to de Donder gauge}
\label{HSPdeDonder}
In addition to the nonlocality just noted, there are some other puzzles in the dictionary~(\ref{gkpwlc}).  In a covariant form, the operator on the RHS would be a spin-$S$ current bilinear, of dimension $\Delta = S+1$.  The normalizable and nonnormalizable bulk modes would then have the behaviors $z^{S+1}$ and $z^{2-S}$.  The exponents~(\ref{lcexp}) correspond to $\Delta = 1$.  This is natural if we include the $(\partial_{x_1^-} + \partial_{x_2^-})^{-S}$ in the dimension of the operator, but we would like to understand in detail how the light-cone and covariant behaviors are related, and why the even spins have the alternate-quantization~$z^1$ behavior, while the odd spins have the normal-quantization $z^2$ behavior.  Finally, we would like to understand why the boundary behavior of $h_{x(2l)}$ is related to the current $j_{-(2l)}$, while $h_{x(2l+1)}$ is related to $\partial_{ x} j_{-(2l+1)} + \partial _- j_{x-(2l)}$ (the conserved currents are reviewed in the Appendix).
 
To answer these questions, we will work out the GKPW dictionary in de Donder gauge, and then directly transform the asymptotics to light-cone gauge.  In the notation of Metsaev~\cite{Metsaev:1999ui}, de Donder gauge is given by
\be
\bar{\alpha} D |\Phi\rangle = 0 \,.
\ee
In Ref.~\cite{Mikhailov:2002bp}, it is shown that there is a residual gauge symmetry that allows us to restrict to traceless fields
\be
\bar{\alpha}^2 |\Phi\rangle = 0 \, .
\ee
Then the higher spin field equation~\ref{HSEoMNoGauge} can be written
\be
\begin{split}
\Big ( & \hat{\partial}_A^2 + d \hat{\partial}_z + 2 M_{zM} \hat{\partial}_M - \alpha^2_{M} \bar{\alpha}^2_N + \left ( \alpha_z \bar{\alpha}_z \right )^2 \\ &  - (d + 2S - 2) \alpha_z \bar{\alpha}_z - S^2 + (4-d) S + 2d-4 \Big )| \Phi \rangle= 0 \, ,
\end{split}
\ee
where $M$ and $N$ run over the Poincar\'{e} indices.  We are interested in the dictionary for $z\to0$, so we may drop the $\hat{\partial}_M$ terms, which are higher order in $z$.  

Consider first $|\Phi\rangle$ with only Poincar\'{e} indices, and traceless.  The equation of motion is
\be
\left ( \hat{\partial}_z^2 + d \hat{\partial}_z - S^2 + (4-d)S + 2d-4 \right ) h_{M_1 \dots M_S} = 0 \, ,
\ee
with solutions
\be
h_{M_1 \dots M_S} \sim z^{2-S} a_{M_1  \dots M_S}(x) + z^{S+d-2} b_{M_1 \dots M_S}(x) \, .
\ee
Next, the equation of motion for $|\Phi\rangle$ with exactly one component in the $z$ direction is
\be
\left ( \hat{\partial}_z^2 + d \hat{\partial}_z - S^2 + (2-d) S + d-1 \right ) h_{M_1 \dots M_{S-1} z} = 0 \, ,
\ee
with solutions
\be
h_{M_1 \dots M_{S-1} z} \sim z^{1-S} a_{M_1  \dots M_{S-1} z}(x) + z^{S+d-1} b_{M_1 \dots M_{S-1} z}(x) \, .
\ee

Finally, consider the de Donder gauge condition 
\be
 \bar{\alpha}_M \hat{\partial}_M  \Phi  =   - \bar{\alpha}_z \left ( \hat{\partial}_z + 1 - d -S \right)  \Phi \, .
\ee
For the normalizable mode $b_{A_1\dots A_S}$, this determines the components with $k$ $z$ indices iteratively in terms of the divergence of those with one fewer $z$ index, except at $k=1$, where we just get $\partial_{M_1} b_{M_1 \dots M_{S-1} z} = 0$.  Each additional $z$ index brings one additional power of $z$ (from the $\hat{\partial}_M$), so the normalizable terms go as $z^{S+d+k-2}$.  Also, the $\bar{\alpha}^2$ condition relates each Poincar\'{e} trace to a component with two additional $z$ indices, so it is smaller than the traceless part by order $z^2$.  Based on conservation and tracelessness, we identify the holographic dictionary 
\be
\label{NormToCurrent}
b_{M_1 \dots M_S} \propto j_{M_1\dots M_S} \, .
\ee

We now transform to light-cone gauge.  The transformation is
\be
\Phi^\prime  = \Phi - \alpha^A D_A \Lambda = \Phi -\left ( \alpha \hat{\partial} + (S-1) \alpha_z - \alpha^2 \bar{\alpha}_z\right ) \Lambda \, .
\ee
On the LHS the $-$ components vanish, so there are no terms with $\alpha_+$.  On the right, we expand in
\be
\Phi  = \sum_{\ell = 0}^S \alpha^\ell_+ \Phi_\ell  \,, \quad \Lambda  = \sum_{\ell = 0}^S \alpha^\ell_+ \Lambda_\ell  \, .
\ee
Then the gauge transformation becomes
\be
\left ( \hat{\partial}_- + 2 \alpha_- \bar{\alpha}_z\right )  \Lambda_\ell  = \Phi_{\ell + 1} - \left( \alpha_- \hat{\partial}_+ + \alpha_x \hat{\partial}_x + \alpha_z \left ( \hat{\partial}_z + S - 1 - \alpha_z \bar{\alpha}_z \right ) - \alpha_x^2 \bar{\alpha_z} \right ) \Lambda_{\ell + 1}  \, ,
\ee 
along with
\be
 \Phi^\prime  = \Phi_{0} - \left( \alpha_- \hat{\partial}_+ + \alpha_x \hat{\partial}_x + \alpha_z \left ( \hat{\partial}_z + S - 1 - \alpha_z \bar{\alpha}_z \right ) - \alpha_x^2 \bar{\alpha_z} \right ) \Lambda_{0}  \, ,
\ee
where $x$ corresponds to indices that are transverse to both $z$ and the light-cone coordinates.  We can now solve iteratively in $\ell$ to determine the light-cone field.  

For example, for $S=1$ we have
\be
\begin{split}
& \Lambda^{(1)}_0  = \hat{\partial}_-^{-1} \Phi_1^{(1)}
\\ & \Phi^{\prime(1)}  = \Phi_0^{(1)} - \left ( \alpha_- \hat{\partial}_+ + \alpha_x \hat{\partial}_x + \alpha_z \hat{\partial}_z \right ) \hat{\partial}_-^{-1} \Phi_1^{(1)} \,,
\end{split}
\ee
or
\be
h_A = h_A - \hat{\partial}_A \hat{\partial}_-^{-1} h_- \, .
\ee
For the asymptotics of the normalizable mode this is
\be
h_{x} \sim z^{d-1} \left (b_{x} - \partial_{x} \partial^{-1}_- b_-\right ) \,, \quad h_z \sim \left ( 2-d\right ) z^{d-2} \partial_-^{-1} b_- \, , \label{LCS1}
\ee
which are in turn related to the currents Eq.\ \ref{NormToCurrent}.  For example, for $d=3$, we have $h_{x} \propto z^2 (j_x - \partial_-^{-1} \partial_x j_-)$ and $h_z  \propto z \partial_-^{-1} j_-$.   Thus, all the puzzling features noted at the beginning of this subsection are accounted for by the transformation from de Donder gauge to light-cone gauge.

For simplicity, we now focus on the $x,z$ components so we can drop $\alpha_-$ terms, and we set $d=3$.  Then at $S=2$, 
\be
\begin{split}
& \Lambda^{(2)}_1 = \hat{\partial}_-^{-1} h_{--}
\\ & \Lambda^{(2)}_0 = \alpha_x \left ( \hat{\partial}_-^{-1} h_{x-} - \hat{\partial}_-^{-2} \hat{\partial}_x h_{--} \right ) + \alpha_z \left ( \hat{\partial}_-^{-1} h_{z-} - \hat{\partial}_-^{-2} \hat{\partial}_z h_{zz} \right ) \, .
\end{split}
\ee
In evaluating $\Phi^\prime$ we want only terms of order $z^{1,2}$.  Since $\Phi$ is of order $z^{3+k}$, most terms drop out, leaving
\be
\begin{split}
\Phi^{\prime(2)} & \approx \alpha_x \alpha_z \left ( \hat{\partial}^{-1}_- \hat{\partial}_z h_{x-} - 2 \hat{\partial}_-^{-2} \hat{\partial}_x \hat{\partial}_z h_{--} \right ) - \alpha_z^2 \hat{\partial}_-^{-2} (\hat{\partial}_z - 2) \hat{\partial}_z h_{--} + \alpha_x^2 \hat{\partial}_-^{-2} \hat{\partial}_z h_{--}
\\ & \approx 3 z^2 \alpha_x \alpha_z ( \partial_-^{-1} b_{x-} - 2 \partial_-^{-2} \partial_x b_{--} ) - 3 z \alpha_z^2 \partial_-^{-2} b_{--} + 3 z \alpha_x^2 \partial_-^{-2} b_{--} \, .
\end{split}
\ee
This is traceless, as desired.  Again, components with an even number of $x$ indices go as $z$ and components with an odd number of $x$ indices go as $z^2$, and the requisite $\partial^{-2}$ appears.

\subsection{Gauge Invariant Dictionary}
As an alternative to the above construction, we could instead phrase the dictionary in terms of gauge invariant bulk objects.  The proper objects are the spin-$S$ Weyl curvatures, which generalize the notion of the spin-$1$ field strength and the spin-$2$ Weyl tensor.  The spin-$S$ Weyl curvature is a traceless $2S$-index tensor whose indices group into $S$ pairs, which are anti-symmetric under exchange of the elements of a pair and symmetric under exchange of pairs:
\be
C_{A_1 A_2 A_3 A_4 \dots A_{2S-1} A_{2S}} = - C_{A_2 A_1 A_3 A_4 \dots A_{2S-1} A_{2S}} = C_{A_3 A_4 A_1 A_2  \dots A_{2S-1} A_{2S}} \, .
\ee
Like their lower spin versions, they possess a Bianchi identity
\be
D_{(A} C_{B C ) A_3 A_4 \dots A_{2S-1} A_{2S}} = 0 
\ee
for AdS covariant derivative $D_{A}$ and antisymmetrization on the coordinates $A,B,C$.  This, combined with their tracelessness, gives the conservation equation
\be
D^{B} C_{ B A_2 A_3 A_4 \dots A_{2S-1} A_{2S}} = 0 \, .
\ee
At linearized level, they are constructed out of $S$ covariant derivatives of the spin-$S$ metric field.  This feature, combined with the properties above, uniquely fixes the Weyl curvature up to an overall constant, and so we may schematically write
\be
C_{A_1 A_2 A_3 A_4 \dots A_{2S-1} A_{2S}} \propto D_{A_1} D_{A_3} \dots D_{A_{2S-1}} h_{A_2 A_4 \dots A_{2S}} + \mathrm{permutations} - \mathrm{traces} \, ,
\ee
where every permutation of an $A_i$ for odd $i$ with one for even $i$ gives a minus sign.  Generalizing from the lower spin dictionary, we then expect the GKPW dictionary
\be
j_{M_1 M_2 \dots M_S} \sim \lim_{z\to 0} z^{-(S+d-2)} C_{M_1 z M_2 z \dots M_S z} \, .
\label{WeylGKPW}
\ee
The necessary scaling in $z$ can be easily determined from the considerations in \S\ref{HSPdeDonder} using either de Donder or light-cone gauge.   

The Weyl curvature component on the right-hand side obeys both Poincar\'{e} conservation and Poincar\'{e} tracelessness:
\bea
\tensor{C}{^{M}}{_{ z M z M_3 z \dots M_S z}} & = & 0 \, ,\nonumber \\ 
\partial^M C_{MzM_2 \dots M_S z} & = & 0 \, .
\eea
These properties result from the bulk conservation and tracelessness of the Weyl curvature, combined with the vanishing of any component with $S+1$ or more $z$ indices.  This assures that the boundary current obtained from Eq.~(\ref{WeylGKPW}) is automatically traceless and conserved.

In light-cone gauge, the simplest current component to write down is the all minus current component, for which all but one term vanishes due to the gauge constraint:
\be
j_{-(S)} \sim \lim_{z\to 0} z^{-(S+d-2)} C_{- z - z \dots - z} \propto  \lim_{z\to 0} z^{-(d-2)} \left ( \partial_- \right)^S h_{z(S)} \, . \label{allMinusGKPW}
\ee
This has the same form schematic form as the $j_-$ current discussed under Eq.~(\ref{LCS1}).  Note also that although the light-cone higher spin fields may be nonlocal in $\partial_-$, the gauge invariant curvatures are local, as expected.  For example, in Eq.~(\ref{allMinusGKPW}) the factor of $(\partial_-)^S$ exactly cancels the $(\partial_-)^{-S}$ in $h_{z(S)}$ found in Eq.~(\ref{gkpwlc}) and Eq.~(\ref{gkpwlc2}).  
\sect{Discussion}
\label{Conclusion}

We have shown that the RG equation for a cutoff on $z$ in the light-cone theory is equivalent to the corresponding higher spin bulk field equation.  If we treat the covariant path integral~(\ref{covrg}) in the same way, then the collective field equation again becomes the RG equation,
\be
(P/2+i\partial_r)^2 (P/2-i\partial_r)^2 \chi(P,r) = 0\,.
\ee
We have overachieved, obtaining a fourth order equation, but we have not found a way to relate this to the bulk field equation.  It is not clear why the light cone should play such an essential role.  

It is not clear how to extend this approach to obtain tree level interactions in the bulk.  If we include the interactions induced by the nonlinearities~(\ref{pinonlin}), there is no locality in $z$.  Of course, in the Vasiliev theory this is achieved only by adding auxiliary parameters.  Perhaps this can be achieved by combining our cutoff with the formalism of Refs.~\cite{Douglas:2010rc,Leigh:2014tza}.  Thus far, however, it must be said that the RG approach has added only limited value: we have just given an RG interpretation to the work of Refs.~\cite{Koch:2010cy,Jevicki:2011ss} 

Finally we make a few remarks about bulk interactions in the precursor approach.  As emphasized already in the early work~\cite{BDHM}, the leading precursor formulae such as that of Ref.~\cite{Koch:2010cy}, Eq.~(\ref{phid3}, \ref{chid3}), must receive corrections.  The bulk field as constructed in leading order exactly satisfies a linear wave equation, exhibiting no bulk interactions.  The $1/N$ corrections was developed in Refs.~\cite{Kabat:2011rz,Heemskerk:2012mn}, using on local commutativity in the bulk or alternatively the bulk field equations.  Either will be challenging here: we are dealing with gauge fields, and their gauge-fixed commutators will not be fully local, and the appropriate gauge fixing of the Vasiliev interaction is not clear.  

We note also an apparently independent source of $1/N$ corrections.  The light-cone GKPW dictionary~(\ref{gkpwlc}) must be local (up to gauge fixing issues), and this requires inclusion of the momentum ranges $p_1^+ > P^+$ and $p_1^+ < 0$, which are omitted in the linearized precursor.  

\section*{Acknowledgements}

We thank Cheng Peng for many discussions, and early collaboration on the material of \S\ref{HRG}.  We also thank Albion Lawrence, Sung-Sik Lee,  Rob Leigh, and Leo Pando-Zayas for discussions.  This work was supported by NSF grants PHY07-57035, PHY11-25915, and PHY13-16748.

\appendix
\sect{Traceless Conserved Currents}
\label{TracelessCurrents}
The GKPW dictionary maps the bulk invariant Weyl curvatures and the boundary traceless conserved currents.  In order to check that these objects do indeed agree, we need to determine the traceless conserved currents directly from the boundary.  This result has been written down in several places and in several forms, first in \cite{Anselmi:1999bb}, but also in \cite{Das:2003vw, Mikhailov:2002bp, Giombi:2009wh}.  Here we re-derive the result in a manner similar to \cite{Giombi:2009wh} in order to obtain yet another form for the currents. 

Most generically, a current of spin $S$ is a symmetric $S$-index object made up of $S$ derivatives of $\phi_1^a(x) \phi^a_2(x)$.  For real fields, the two fields are not distinct, but we will label them with a $1$ or $2$ so that it is clear whether a derivative acts on the first ($\partial^1_M$) or second ($\partial^2_M$) field.  For complex fields, $\phi_1^a(x)$ and $\phi_2^a(x)$ are complex conjugates of each other.  The free field equations of motion give that $\partial^1 \cdot \partial^1 = \partial^2 \cdot \partial^2 = 0$.  This leaves three structures that the current may be constructed out of: $\partial^1_M$, $\partial^2_M$, or $\eta_{M N} \partial^1 \cdot \partial^2$.  To parameterize this, contract the current with a polarization vector $\epsilon^M$, then write the result as a function $f_S(u,v,w^2)$:
\be
j_{M(S)} (\epsilon^M )^S =   f_S(\epsilon \bracedot (\partial^1+\partial^2), \epsilon \bracedot (\partial^1-\partial^2), \epsilon^2 \partial^1 \bracedot \partial^2)  \phi^a_1(x) \phi^a_2(x) \equiv f_S(u,v,w^2)  \phi^a_1(x) \phi^a_2(x) \, .
\ee
This $f_{S}(u,v,w^2)$ has no direct relation to the $f_{S,J}(u,v)$ of \S\ref{MatchRGHS}.  To assure the current is conserved, traceless, and spin $S$, we require that it obeys the following:
\begin{enumerate}
\item Conservation:
\be
\left ( \partial^1 + \partial^2 \right ) \bracedot \partial_{\epsilon} \left ( j_{M(S)} (\epsilon^M )^S \right ) = 0
\ee
\item Tracelessness:
\be
\partial_\epsilon \bracedot \partial_\epsilon \left ( j_{M(S)} (\epsilon^M )^S \right ) = 0
\ee
\item Definite spin:
\be
\epsilon \bracedot \partial_\epsilon \left ( j_{M(S)} (\epsilon^M )^S \right ) = S \left ( j_{M(S)} (\epsilon^M )^S \right )
\ee 
\end{enumerate}
We can convert these requirements to constraints on $f_S$ by making use of the free field equations of motion.  In this form the constraints are
\bea
& 1)& \ f_{S;u} + u f_{S;w^2} = 0 \, ,\nonumber
\\ & 2)&\ f_{S;uu}  -f_{S;vv} + d f_{S;w^2} +  2u f_{S;uw^2} +  2v f_{S;vw^2} + 2 w^2 f_{S;w^2 w^2} = 0 \, ,\nonumber
\\ & 3)&\ u f_{S;u} + v f_{S;v} + 2 w^2 f_{S;w^2} =S f_S \, ,
\eea
where a variable after the semicolon denotes differentiation with respect to that variable.  We may then expand term by term in the `trace' variable $w^2$
\be
f_S (x,y,w^2) = \sum_{n=0}^{\left \lfloor S/2 \right \rfloor} f_{S,n} (x,y) \left ( w^2 \right)^{n} \, ,
\ee
where the upper bound on the sum is included since we seek a polynomial in $\epsilon$ of order $S$.  Finally, it is convenient to make the change of coordinates
\be
k=\frac{v}{u} \, , \quad \quad \quad \quad \ell=u \, .
\ee
The three constraints are then
\bea
&1) & \ - \frac{k}{\ell} f_{S,n;k} + f_{S,n;\ell} + \ell (n+1) f_{S,n+1} = 0 \, , \nonumber
\\ & 2)&\  \frac{2k}{\ell^2} f_{S,n;k} + \frac{k^2}{\ell^2} f_{S,n;kk} -  \frac{2k}{\ell} f_{S,n;k\ell}  + f_{S,n;\ell\ell} \nonumber \\ & & \ \quad -  \frac{1}{\ell^2} f_{S,n;kk} + (n+1) \left [ (2n+d) f_{S,n+1} + 2 \ell f_{S,n+1;\ell}  \right ] = 0 \, ,\nonumber
\\ & 3)&\  \ell f_{S,n;\ell} = (S-2n) f_{S,n} \, . \label{TCCConstraints}
\eea
The third equation gives the simple result
\be
f_{S,n}(k,\ell) = K_{S,n}(k) \ell^{S-2n} \, ,
\ee
while the first equation gives $f_{S,n+1}$ in terms of $f_{S,n}$.  Plugging both of these into the second equation gives the differential equation
\be
\left ( 1 - k^2 \right ) K^{\prime \prime}_{S,n} - \left ( 2n + d -2\right ) k K^\prime_{S,n} + (S-2n) (  S + d - 3) K_{S,n} = 0 \, .
\ee
The polynomial solution to this differential equation is related to the associated Legendre polynomial:
\be
K_{S,n}(k) = c_{S,n} (1-k^2)^{-(d-4+2n)/4} P_{S-n+d/2-2}^{n+d/2-2}(k) \, .
\ee
We can then use the first constraint on $f_{S,n}$ to relate the leading constants $c_{S,n}$ for equal $S$ but different $n$.  This gives the complete result
\be
f_{S,n}(k,\ell) = c_{S,0} \frac{(-1)^n}{n!} (1-k^2)^{-(d-4+2n)/4} P_{S-n+d/2-2}^{n+d/2-2}(k) \ell^{S-2n} \, .
\ee
We may put this in terms of the original variables to get the form
\bea
j_{M(S)} (\epsilon^M )^S = c_{S,0} \Bigg \{ & & \sum_{n=0}^{\lfloor S/2 \rfloor} \frac{(-1)^n}{n!}\left [1-\left (\frac{\epsilon \cdot \left (\partial^1 - \partial^2\right )}{\epsilon \cdot\left (\partial^1 + \partial^2 \right )} \right ) ^2 \right ]^{-(d-4+2n)/4} \nonumber\\ & & \times P_{S-n+d/2-2}^{n+d/2-2}\left ( \frac{\epsilon \cdot \left (\partial^1 - \partial^2\right )}{\epsilon \cdot\left (\partial^1 + \partial^2 \right )} \right ) \left ( \epsilon \cdot\left (\partial^1 + \partial^2 \right ) \right )^{S-2n}  \left ( \epsilon^2 \partial^1 \cdot \partial^2 \right )^n \Bigg \} \phi_1^a(x) \phi_2^a(x) \nonumber \\ \label{TCCurrents}
\eea
Though this may look non-local, it is not: the solution to $K_{S,n}$ is a polynomial of maximum degree $S-2n$, and so all factors of $\epsilon \cdot\left (\partial^1 + \partial^2 \right )$ in the denominator are cancelled.
\subsection{Example Currents}
Unpacking a general component of the currents in Eq.~(\ref{TCCurrents}) requires $S$ derivatives with respect to $\epsilon$ and is not straightforward to write down.  Here we consider a couple of the simpler components in order to compare to the bulk results.  

The easiest example is the component of the current with all minus light-cone indices, $j_{-(S)}$.  The trace term in  Eq.~(\ref{TCCurrents}) does not contribute to this component, so we get just
\be
j_{-(S)} = c_{S,0} \left [ 1 -\left ( \frac{\partial^1_- - \partial^2_-}{\partial^1_- + \partial^2_-} \right )^2\right ]^{-(d-4)/4} P_{S+d/2-2}^{d/2-2} \left ( \frac{\partial^1_- - \partial^2_-}{\partial^1_- + \partial^2_-} \right ) \left ( \partial^1_- + \partial^2_- \right )^S \phi_1^a(x) \phi_2^a(x) \, . \label{allMinusCurrent}
\ee
We want to check the relation 
\be
j_{-(S)} \propto \lim_{z\to0} z^{-(d-2)} (\partial_-)^S h_{z(S)} \, .
\ee
arising from Eq.~(\ref{allMinusGKPW}). The metric component $h_{z(S)}$ is proportional to the $J=0$ part of the mapping Eq.~(\ref{ftog}), and is
\be
h_{z(S)} = z^{d-2}g_{S,0}(\cos \theta) \phi_1^a(x) \phi_2^a(x) \, .
\ee
With the usual coordinate transform $\partial^1_- + \partial^2_- = \partial_-$ and $\partial^1_- - \partial^2_- = \partial_- \cos \theta$, we may compare Eq.~(\ref{allMinusCurrent}) to Eq.~(\ref{gSJ}) for $J=0$ and see that the results agree.  

To illustrate the process required to compare general current components, consider the component with a single transverse index, $j_{i-(S-1)}$.  The trace term still does not contribute, so we must calculate
\bea
j_{i-(S-1)} =  \frac{\partial}{\partial \epsilon^i} \Bigg \{ & & \!  c_{S,0}\left [1-\left (\frac{\epsilon \cdot \left (\partial^1 - \partial^2\right )}{\epsilon \cdot\left (\partial^1 + \partial^2 \right )} \right ) ^2 \right ]^{-(d-4)/4}  P_{S+d/2-2}^{d/2-2}\left ( \frac{\epsilon \cdot \left (\partial^1 - \partial^2\right )}{\epsilon \cdot\left (\partial^1 + \partial^2 \right )} \right ) \nonumber \\ & & \times \left ( \epsilon \cdot\left (\partial^1 + \partial^2 \right ) \right )^{S}  \phi_1^a(x) \phi_2^a(x) \Bigg \} \Bigg |_{\epsilon^m=\delta^m\!_-} \, .
\eea
It is now convenient to switch back to the language of $K_{S,n}(k)$ and $\ell$.  We may use (1) in Eq.~(\ref{TCCConstraints}) to determine that 
\be
K_{S,n}^\prime(k) = \frac{1}{k} \left [ (S-2n) K_{S,n}(k) + (n+1) K_{S,n+1}(k) \right ] \, .
\ee
So the current component becomes
\be
j_{i-(S-1)} = \left . c_{S,0} \left [S K_{S,0}(k) \ell^{S-1} \frac{\partial \ell}{\partial \epsilon^i} + \frac{1}{k} \left ( S K_{S,0}(k) +  K_{S,1}(k) \right ) \ell^S \frac{\partial k}{\partial \epsilon^i} \phi_1^a(x) \phi_2^a(x) \right ] \right |_{\epsilon^m=\delta^m\!_-}\, .
\ee
From the definitions of $\ell$ and $k$ we have 
\bea
\frac{\partial \ell}{\partial \epsilon^i} &=& (\partial_i^1 + \partial^2_i) \nonumber \\
\frac{\partial k}{\partial \epsilon^i} &=& \frac{1}{\ell}  (\partial_i^1 - \partial^2_i) - \frac{k}{\ell} (\partial_i^1 + \partial^2_i)
\eea
which leads to 
\bea
j_{i-(S-1)} =  c_{S,0} \Bigg [ & & \!  \frac{1}{k} (S K_{S,0}(k) + K_{S,1}(k) )\ell^{S-1} (\partial_i^1 - \partial^2_i)  \nonumber \\ & & -  K_{S,1}(k) \ell^{S-1} (\partial_i^1 + \partial^2_i)  \Bigg ] \phi_1^a(x) \phi_2^a(x)\Bigg |_{\epsilon^m=\delta^m\!_-}\, .
\eea
First, we want to rewrite
\be
\partial^1_i - \partial^2_i = \left [ k (\partial_i^1 + \partial^2_i) + \sqrt{1-k^2} \frac{\partial^2_- \partial^1_i - \partial^1_- \partial^2_i}{\sqrt{\partial^1_- \partial^2_-}}\right ] \Bigg |_{\epsilon^m=\delta^m\!_-} \, ,
\ee
since the derivative of the parameter $\vec{z}$ corresponds to $(\partial^2_- \partial^1_i - \partial^1_- \partial^2_i)/\sqrt{\partial^1_- \partial^2_-}$ as a result of Eq.~(\ref{defvecz}).  This then gives
\bea
j_{i-(S-1)} =  c_{S,0} \Bigg [ & & \! \frac{\sqrt{1-k^2}}{k} (S K_{S,0}(k) + K_{S,1}(k) )\ell^{S-1}\frac{\partial^2_- \partial^1_i - \partial^1_- \partial^2_i}{\sqrt{\partial^1_- \partial^2_-}} \nonumber \\ & & + S K_{S,0}(k) \ell^{S-1} (\partial_i^1 + \partial_i^2) \Bigg ] \phi_1^a(x) \phi_2^a(x)\Bigg |_{\epsilon^m=\delta^m\!_-}\, .
\eea
This may be simplified via associated Legendre polynomial recursion relations.  We have
\be
S K_{S,0}(k) + K_{S,1}(k) =  ( 1- k^2)^{-(d-4)/4} \left [ S P_{S+d/2-2}^{d/2-2}(k) + \frac{1}{\sqrt{1-k^2}} P^{d/2-1}_{S+d/2-3}(k) \right ] \, .
\ee
The associated Legendre polynomials obey
\be
 (l - m) P^m_l(x)  + \frac{P^{m+1}_{l-1}(x)}{\sqrt{1-x^2}} = \frac{x}{\sqrt{1-x^2}} P^{m+1}_l(x) \, ,
\ee
which allows us to write
\be
\frac{1}{k} (S K_{S,0}(k) + K_{S,1}(k)) = (1-k^2)^{-(d-2)/4} P^{d/2-1}_{S+d/2-2}(k) \, .
\ee
In order to compare to the bulk, we want to write the current in the form
\bea
j_{i-(S-1)} =  c_{S,0} (\partial_-)^{S-1} \Bigg [ & & \! (\sin \theta)^{-(d-4)/4} P^{d/2-1}_{S+d/2-2}(\cos \theta) \frac{\partial^2_- \partial^1_i - \partial^1_- \partial^2_i}{\sqrt{\partial^1_- \partial^2_-}} \nonumber \\ & & + S K_{S,0}(\cos \theta) (\partial^i_1 + \partial^i_2)  \Bigg ]  \phi_1^a(x) \phi_2^a(x) \, .
\eea
Using Eq.~(\ref{gSJ}), this can then be written
\be
j_{i-(S-1)} =  S c_{S,0} (\partial_-)^{S-1} \Bigg [  \frac{S+d-3}{d-2} g_{S,1}(\cos \theta) \frac{\partial_2^+ \partial_1^i - \partial_1^+ \partial_2^i}{\sqrt{\partial_1^+ \partial_2^+}}  +  g_{S,0}(\cos \theta) \partial^i_x  \Bigg ]  \phi_1^a(x) \phi_2^a(x) \, . \label{mostlyMinusCurrent}
\ee
Our bulk-to-boundary mapping gives
\bea
h_{z(S)} &=& z^{d-2}g_{S,0}(\cos \theta) \phi_1^a(x) \phi_2^a(x) \, , \nonumber \\
h_{iz(S-1)} &=& z^{d-1} g_{S,1}(\cos \theta) \frac{\partial^2_- \partial^1_i - \partial^1_- \partial^2_i}{\sqrt{\partial^1_- \partial^2_-}} \phi_1^a(x) \phi_2^a(x) \, .
\eea
If we reverse the relations given under Eq.~(\ref{LCS1}), we see that Eq.~(\ref{mostlyMinusCurrent}) does take the correct general form.  The expected relative coefficient in the bulk between the $g_{S,1}$ and $g_{S,0}$ terms could be determined by a careful consideration of all contributions to the $h_{iz(S-1)}$ term in the Weyl curvature.

Note that for $d=3$, the differential equation for $K_{S,n}(k)$ is that of a Chebyshev polynomial of the first kind for $n=0$ and that of a Chebyshev polynomial of the second kind for $n=1$.  The above check then immediately works for $d=3$, and the field components in Eq.~(\ref{gkpwlc}) and Eq.~(\ref{gkpwlc2}) will lead to the correct boundary conserved currents.

\end{document}